\documentclass[reprint,amsmath,amssymb,aps,]{revtex4-2}

\usepackage{graphicx}
\usepackage{dcolumn}
\usepackage{longtable}
\usepackage{bm}
\usepackage{mathtools}
\usepackage{braket}
\usepackage[]{hyperref}
\usepackage{xcolor} 
\usepackage[caption=false]{subfig}
\usepackage{amsmath}

\begin{document}

\title{Quantum interface for telecom frequency conversion based on \\
	diamond-type atomic ensembles}

\author{Po-Han Tseng$^{1}$, Ling-Chun Chen$^{1,2}$, Jiun-Shiuan Shiu$^{1,2}$, and Yong-Fan Chen$^{1,2}$}

\email{yfchen@mail.ncku.edu.tw}

\affiliation{$^1$Department of Physics, National Cheng Kung University, Tainan 70101, Taiwan \\
	$^2$Center for Quantum Frontiers of Research $\&$ Technology, Tainan 70101, Taiwan}

\date{January 18, 2024}


\begin{abstract}

In a fiber-based quantum network, utilizing the telecom band is crucial for long-distance quantum information (QI) transmission between quantum nodes. However, the near-infrared wavelength is identified as optimal for processing and storing QI through alkaline atoms. Efficiently bridging the frequency gap between atomic quantum devices and telecom fibers while maintaining QI carried by photons is a challenge addressed by quantum frequency conversion (QFC) as a pivotal quantum interface. This study explores a telecom-band QFC mechanism using diamond-type four-wave mixing (FWM) with rubidium energy levels. The mechanism converts photons between the near-infrared wavelength of 795 nm and the telecom band of 1367 or 1529 nm. Applying the Heisenberg-Langevin approach, we optimize conversion efficiency (CE) across varying optical depths while considering quantum noises and present corresponding experimental parameters. Unlike previous works neglecting the applied field absorption loss, our results are more relevant to practical scenarios. Moreover, by employing the reduced-density-operator theory, we demonstrate that this diamond-type FWM scheme maintains quantum characteristics with high fidelity, unaffected by vacuum field noise, enabling high-purity QFC. Another significant contribution lies in examining how this scheme impacts QI encoded in photon-number, path, and polarization degrees of freedom. These encoded qubits exhibit remarkable entanglement retention under sufficiently high CE. In the case of perfect CE, the scheme can achieve unity fidelity. This comprehensive exploration provides theoretical support for the application of the diamond-type QFC scheme based on atomic ensembles in quantum networks, laying the essential groundwork for advancing the scheme in distributed quantum computing and long-distance quantum communication.

\end{abstract}


\maketitle


\newcommand{\FigOne}{
    \begin{figure}[t]
    \centering
    \includegraphics[width = 8.0 cm]{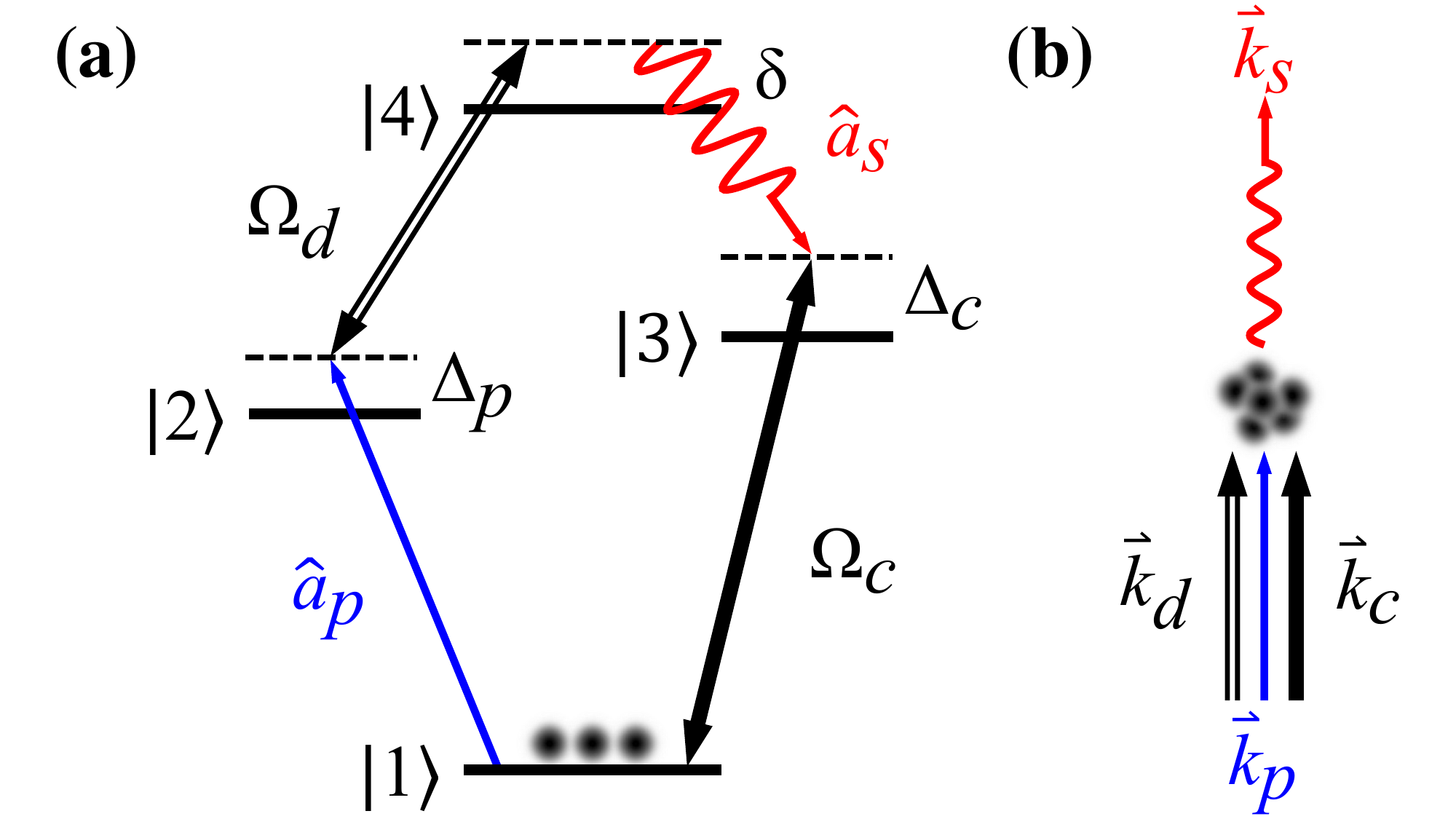}
    \caption{
Diamond-type atomic ensemble QFC system. (a) Energy level diagram illustrating the QFC scheme and the corresponding transitions for the four participating fields. Note that no degenerate states of the energy levels are considered here. However, in Sec. \ref{subsec:conversion efficiency optimization}, we delve into the actual energy level configurations. (b) Schematic diagram illustrating the propagation directions of the participating fields. All light fields propagate in the same direction, nullifying the phase mismatch in the system. The presented scenario involves frequency down-conversion, where the probe field is transformed into the signal field. Conversely, up-conversion proceeds in the opposite manner, converting the signal field back to the probe field.
}
    \label{fig:energy level diagram}
    \end{figure}
}

\newcommand{\FigTwo}{
    \begin{figure}[t]
    \centering
    \includegraphics[width = 8.8 cm]{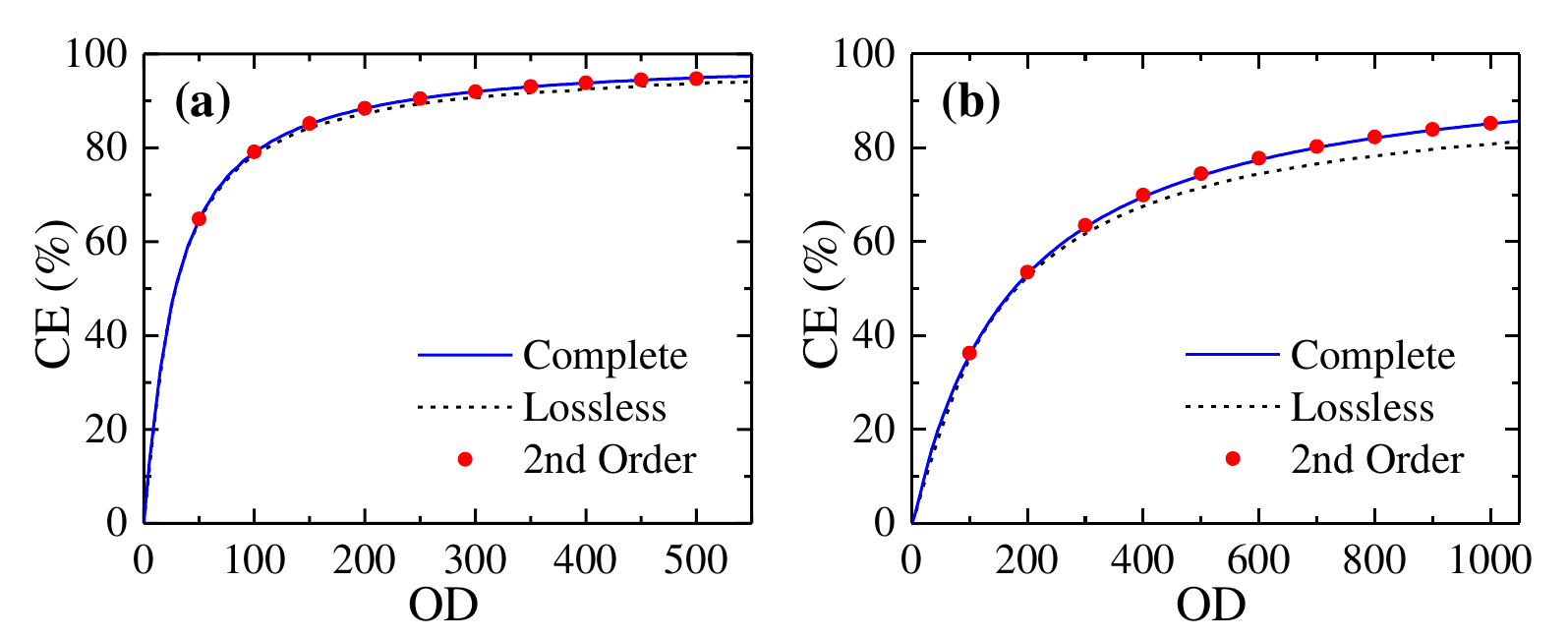}
    \caption{
Optimized CE is presented as a function of OD in the blue solid curve for both (a) telecom E-band QFC scheme and (b) telecom C-band QFC scheme. Here, OD is defined as $\alpha=n\sigma_p L\times\frac{1}{2}$, taking into account the transition coefficient of the selected transition scheme. The red solid circles represent the CE calculated using the Magnus expansion up to the second order. The black dotted curves show the CE curves for the model without considering the coupling field absorption loss. For the optimized five parameters and spontaneous emission rates of the two QFC schemes, please refer to Table \ref{table:optimized parameters}.
}
    \label{fig:Optimized Curve}
    \end{figure}
}

\newcommand{\FigThree}{
    \begin{figure}[t]
    \centering
    \includegraphics[width = 8.8 cm]{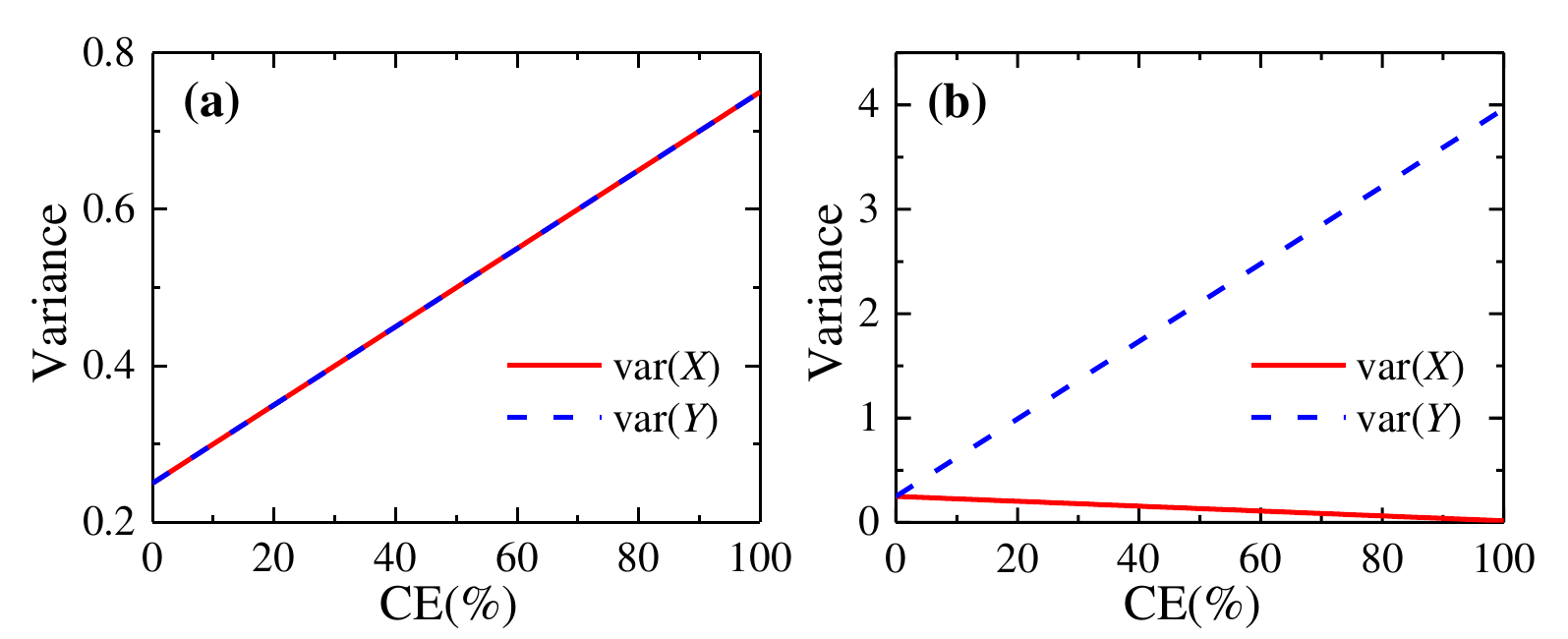}
    \caption{
Quadrature variances are analyzed as a function of CE for the input probe field in both (a) the single-photon Fock state and (b) the 6-dB squeezed coherent state. The red and blue curves represent the quadrature variances $X$ and $Y$ of the output signal field, respectively. In scenario (b), the phase of the signal field is nullified by a phase shifter, and the input field is squeezed with $\phi=0$. Notably, the relative squeezing strength is defined as $P=-10\log_{10}\frac{\sqrt{{\rm var}(X_i)}}{1/2}$, where $P$ measured in dB, and ${\rm var}(X_i)=\frac{1}{4}e^{-2r}$.
}
    \label{fig:quadrature variance}
    \end{figure}
}

\newcommand{\FigFour}{
	\begin{figure}[t]
		\centering
		\includegraphics[width = 8.5 cm]{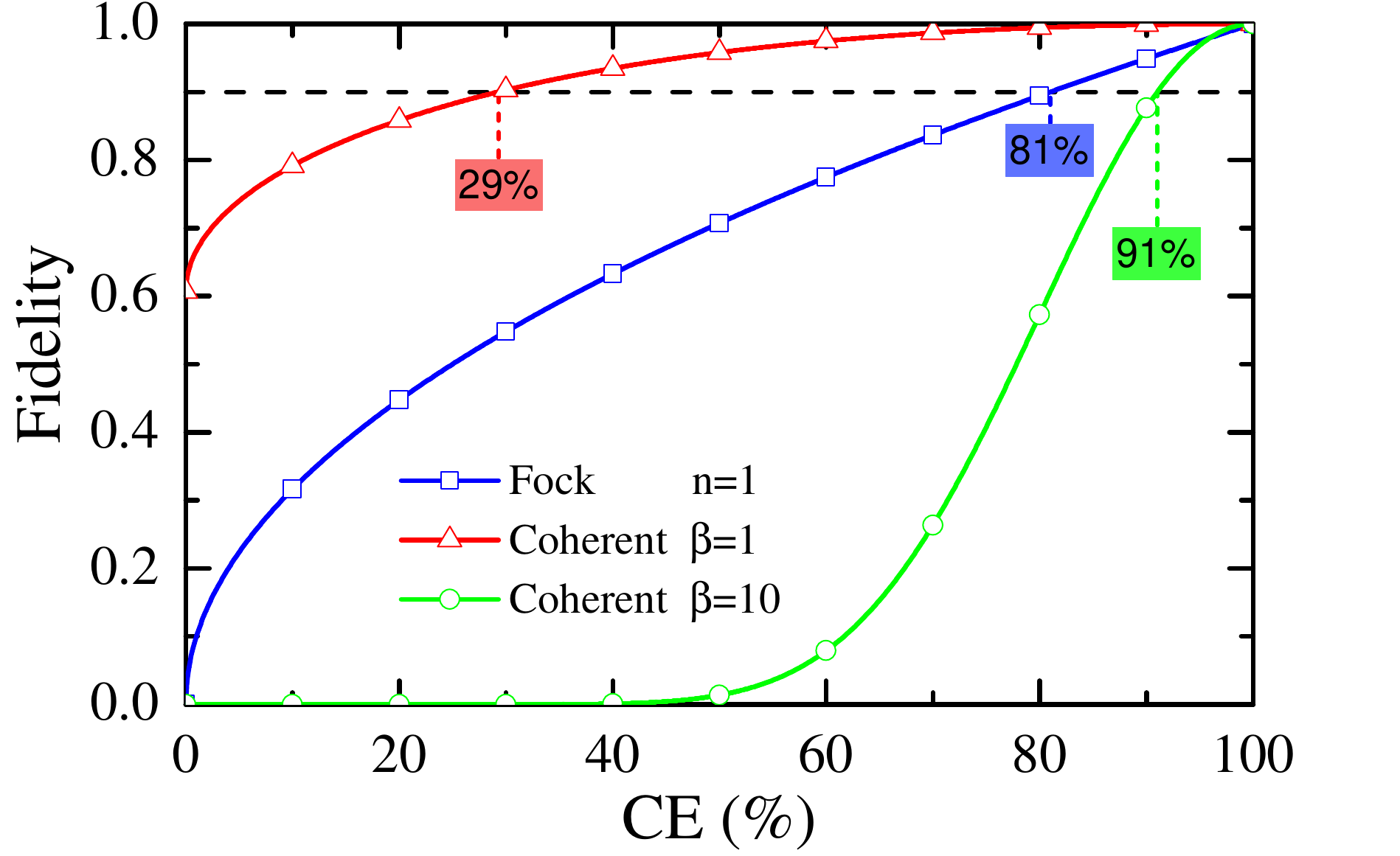}
		\caption{
Fidelity between the input field and the converted field as a function of CE. The blue, red, and green lines correspond to the theoretical curves for the single-photon Fock, single-photon coherent, and 10-photon coherent input states, respectively. The phase of the converted field is eliminated by a phase shifter for both coherent input cases.
}
    \label{fig:fidelity}
	\end{figure}
}

\newcommand{\FigFive}{
	\begin{figure}[t]
		\centering
		\includegraphics[width = 8.5 cm]{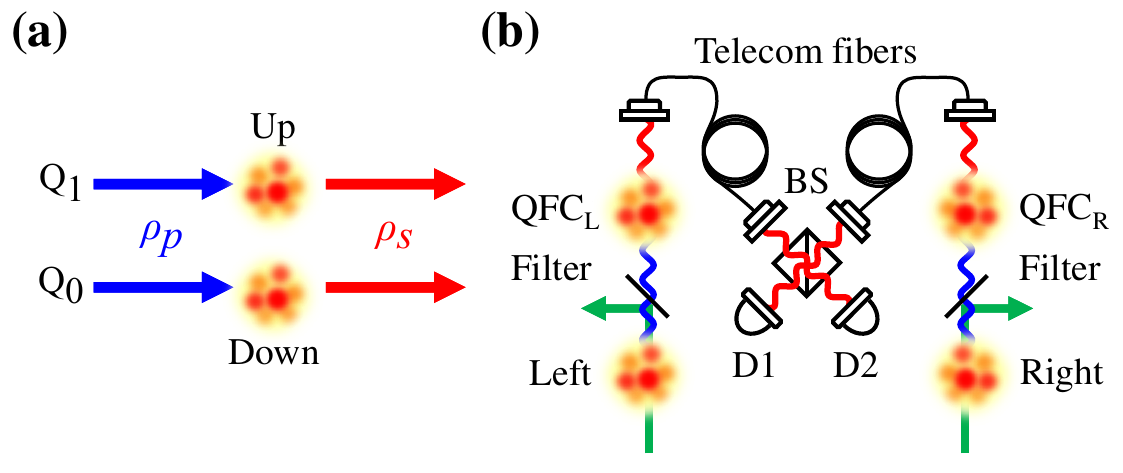}
		\caption{
QFC scheme for path-encoded qubits integrated with the DLCZ protocol. (a) Schematic diagram depicting the frequency down-conversion of a path-encoded qubit. The conversion scheme comprises two spatially separated atomic ensembles. (b) Schematic diagram demonstrating the integration of the telecom-band QFC with the DLCZ protocol. The left and right rubidium atomic ensembles are illuminated by the writing beams (green arrows). The generated Stokes fields (blue wavy lines) are down-converted into the telecom band through the diamond-type QFC scheme after the writing beams are filtered. The converted fields (red wavy lines) are coupled to the telecom fibers for long-distance transmission. The transmitted fields interfere at a 50/50 beam splitter (BS) and are detected by two single-photon detectors, D1 and D2, respectively. The left and right ensembles are entangled if only one of the detectors is clicked.
}
    \label{fig:path encoded qubit}
	\end{figure}
}

\newcommand{\FigSix}{
	\begin{figure}[t]
		\centering
		\includegraphics[width = 8.5 cm]{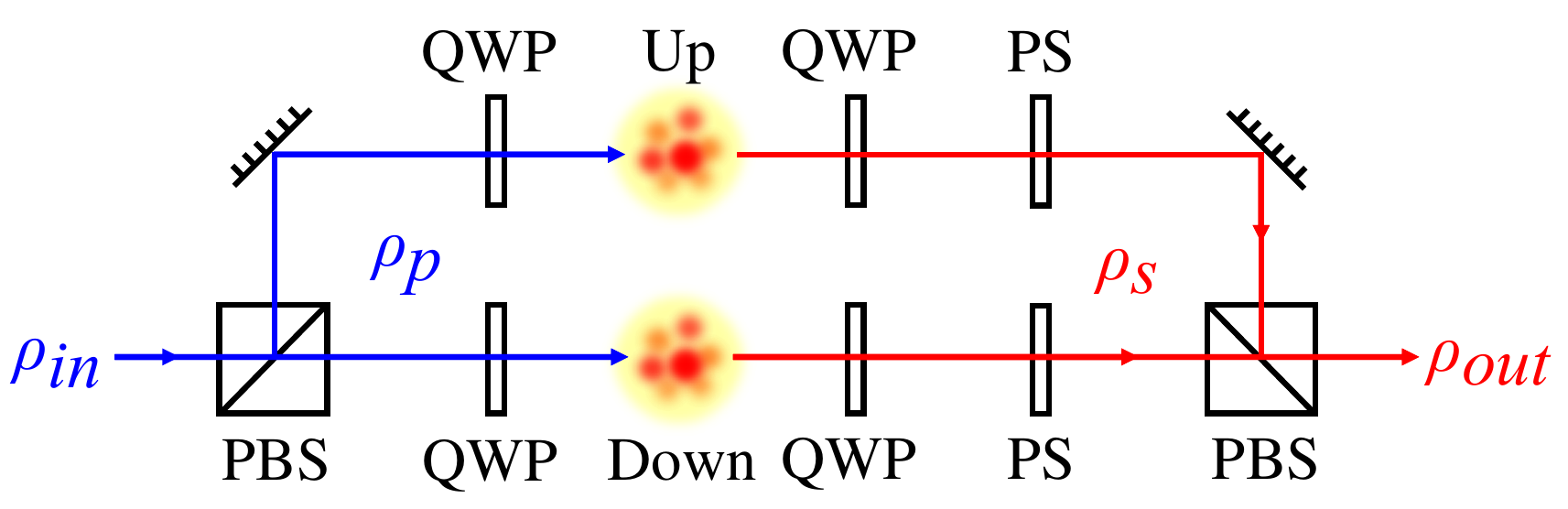}
		\caption{
QFC scheme for a polarization-encoded qubit. Here, $\rho_{in}$ and $\rho_{out}$ denote the input and converted polarization-encoded photons, respectively. The input photon initially passes through a PBS, separating the $\ket{H}$ and $\ket{V}$ components into distinct paths. Two QWPs in separate paths convert both fields to circular polarizations, aligning with the selected QFC schemes. Following the QFCs, another two QWPs, one in each path, revert the fields to horizontal (down) and vertical (up) polarization. Two phase shifters (PSs) are employed to eliminate the phases of $C_U(0)$ and $C_D(0)$, as well as the phase changes induced by the two PBSs. In the final step, the PBS recombines the distinct polarization components back into a single polarization qubit.
}
    \label{fig:polarization encoded qubit}
	\end{figure}
}

\newcommand{\FigSeven}{
	\begin{figure}[t]
		\centering
        \includegraphics[width = 8.6 cm]{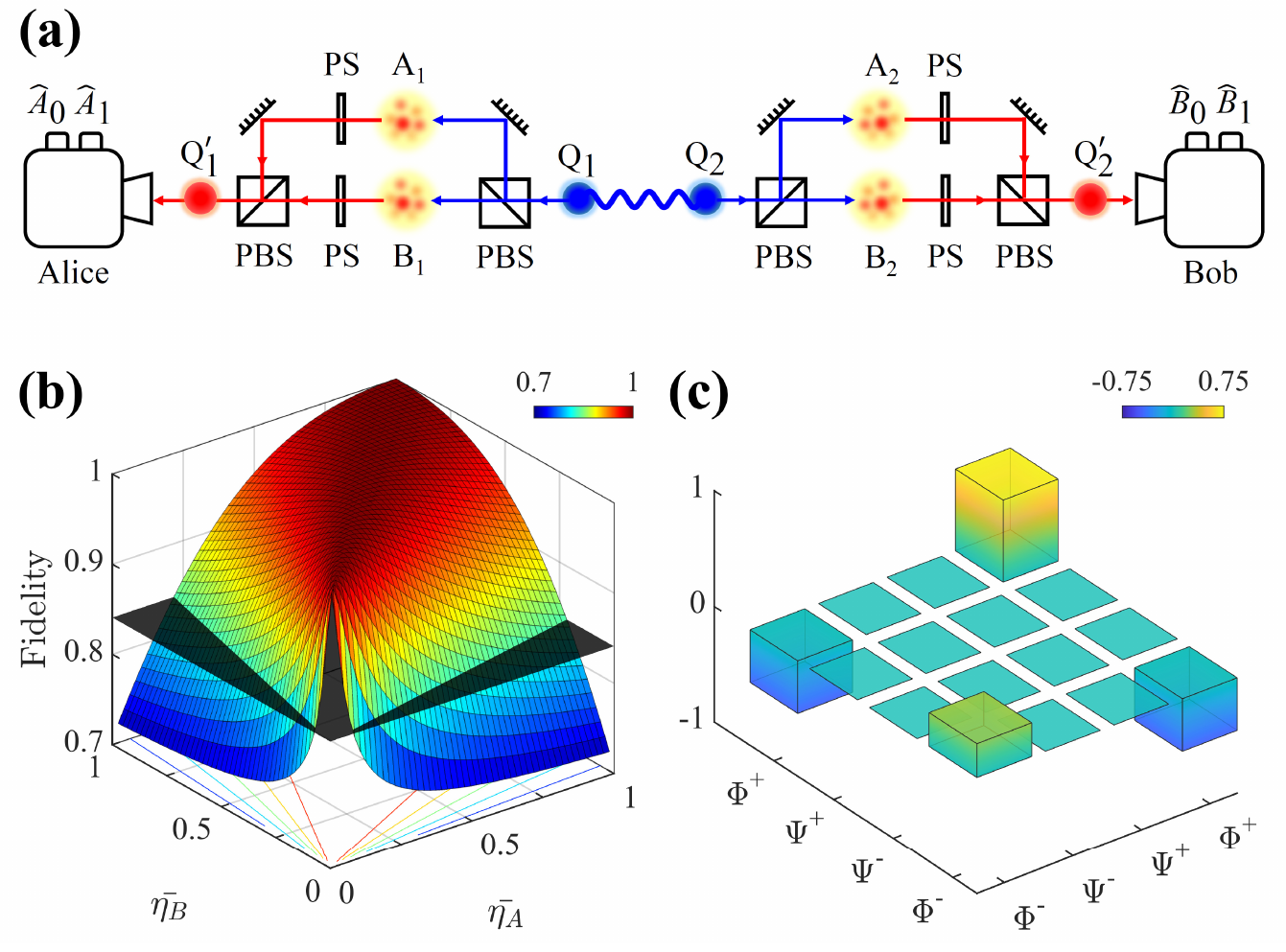}
		\caption{
(a) QFC scheme for a pair of polarization entangled qubits $Q_1$ and $Q_2$. Following the similar setups as in Fig. \ref{fig:polarization encoded qubit} for both qubits, the qubits are converted into another wavelength with the fidelity determined by the four separate QFC setups. In the Bell test, the converted qubits $Q_1'$ and $Q_2'$ are sent to Alice and Bob, respectively. The selected local observables $\hat{A}_0$, $\hat{A}_1$, $\hat{B}_0$, and $\hat{B}_1$ are depicted in Eqs. (\ref{eq:(92)})--(\ref{eq:(95)}). (b) The post-selected fidelity between the input $\ket{\Phi^+}$ state and the converted state as a function of $\bar{\eta}_A$ and $\bar{\eta}_B$. The black plane represents the criteria of fidelity ($F=2^{-1/4}\approx84.1\%$) for the CHSH inequality violation; nonlocality exists for the region with fidelity higher than the criteria. (c) The second converted density operator in Eq. (\ref{eq:(97)}) [the right-hand side intersecting line in Fig. \ref{fig:EPR pairs}(b)] with $F=2^{-1/4}$. The non-zero matrix elements are $-0.455$, $0.707$, $0.293$, and $-0.455$ using the Bell basis representation.
}
    \label{fig:EPR pairs}
	\end{figure}
}

\newcommand{\TableOne}{
    \begin{table*}[t]
        \centering
        \setlength{\tabcolsep}{3.7pt}
        \renewcommand{\arraystretch}{1.5}
		\begin{tabular}{c c c c c c c c c c c c c c c c c c c c c c}
 		\hline\hline
 		 & \multicolumn{21}{c}{Telecom band quantum frequency conversion scheme} \\
 		 \cline{2-22}
 		 & \multicolumn{10}{c}{Telecom E-band (1367 nm)} & & \multicolumn{10}{c}{Telecom C-band (1529 nm)} \\
 		 \cline{2-11}\cline{13-22}
 		OD & \multicolumn{2}{c}{50} & \multicolumn{2}{c}{100} & \multicolumn{2}{c}{150} & \multicolumn{2}{c}{200} & \multicolumn{2}{c}{250} &  & \multicolumn{2}{c}{200} & \multicolumn{2}{c}{400} & \multicolumn{2}{c}{600} & \multicolumn{2}{c}{800} & \multicolumn{2}{c}{1000} \\
 		\hline
 		$\Delta_p$ & 13 & 5 & 25 & 6 & 35 & 8 & 47 & 6 & 59 & 7 &  & 26 & 13 & 49 & 24 & 73 & 33 & 93 & 35 & 116 & 44 \\
 		$\Delta_c$ & -31 & -12 & -54 & -21 & -80 & -31 & -99 & -32 & -123 & -24 &  & -31 & -10 & -64 & -15 & -91 & -7 & -119 & -5 & -154 & -2 \\
 		$\delta$ & 14 & 6 & 26 & 10 & 37 & 14 & 50 & 16 & 62 & 20 &  & 25 & 11 & 48 & 21 & 74 & 26 & 94 & 29 & 116 & 31 \\
 		$\Omega_c$ & 50 & 20 & 90 & 33 & 130 & 46 & 170 & 49 & 210 & 50 &  & 74 & 28 & 145 & 50 & 219 & 50 & 280 & 50 & 350 & 50 \\
 		$\Omega_d$ & 7 & 7 & 13 & 12 & 19 & 17 & 25 & 21 & 31 & 26 &  & 9 & 11 & 16 & 19 & 22 & 29 & 29 & 37 & 36 & 47 \\
 		\hline
 		CE & 64.7 & 63.9 & 79.0 & 77.9 & 85.1 & 83.9 & 88.4 & 86.9 & 90.5 & 89.1 &  & 53.1 & 51.6 & 69.5 & 67.3 & 77.4 & 74.0 & 82.1 & 78.7 & 85.1 & 82.4 \\
 		\hline\hline
		\end{tabular}
		\caption{
The optimized parameter sets for maximizing CEs at different ODs. The five parameters are expressed in units of $\Gamma=2\pi\times 6.063$ MHz, and the CE values are given as percentages. For each OD, two sets of optimized parameters are provided. On the left are those obtained from continuous parameter variation without an upper limit, while on the right are those obtained from a restricted parameter scan where the magnitudes of all parameters are kept under $50\Gamma$. The fine-structure decay rates (and squares of the transition coefficients) are as follows for the telecom E-band QFC scheme \cite{Energy Level and CG Coefficient and Double-lambda QFC Theorem, Alkali Atom Calculator}: $\Gamma_{780}=2\pi\times 6.063$ MHz ($1$), $\Gamma_{795}=2\pi\times 5.745$ MHz ($\frac{1}{2}$), $\Gamma_{1324}=2\pi\times 1.008$ MHz ($\frac{1}{2}$), and $\Gamma_{1367}=2\pi\times 2.087$ MHz ($\frac{1}{2}$). For the telecom C-band QFC scheme \cite{4D3/2 Spontaneous Decay Rate, Alkali Atom Calculator}, the values are $\Gamma_{780}=2\pi\times 6.063$ MHz ($1$), $\Gamma_{795}=2\pi\times 5.745$ MHz ($\frac{1}{2}$), $\Gamma_{1476}=2\pi\times 1.703$ MHz ($\frac{1}{2}$), and $\Gamma_{1529}=2\pi\times 0.315$ MHz ($\frac{1}{5}$).
}
	\label{table:optimized parameters}
    \end{table*}
}


\section{Introduction} \label{sec:introduction}

Quantum networks play a crucial role in enabling distributed quantum computing and quantum communication \cite{The Quantum Internet, Fidelity, Quantum Revolution, Distributed Quantum Computing, Quantum Communication}. In a quantum network, quantum information (QI) undergoes processing \cite{Linear Optical Quantum Computing-1, Linear Optical Quantum Computing-2, Linear Optical Quantum Computing-3, Polarization Encoded Qubits Measurement Based QC, Deterministic Photonic Quantum Computing, Optical Quantum Computing Loop Architecture, Scalable Dual-Loop Circuit QIP} and storage \cite{Optical quantum memory-1, Optical quantum memory Qubit Applications, Optical quantum memory-2, Optical quantum memory-3, Polarization Encoded Qubits Quantum Memory Single Atom, Polarization Encoded Qubits Telecom Quantum Memory, Polarization Encoded Qubits Quantum Memory Atomic Ensemble} within individual quantum nodes. These nodes interconnect through quantum channels, ensuring the transport of quantum states with high fidelity and facilitating entanglement distribution across the network \cite{Polarization Encoded Qubits transmission, Time-Bin Quantum Communication, Cambridge Quantum Network, Quantum Communication QKD}. However, quantum devices may operate at distinct optical frequencies, which might not align with the frequency range of fiber-optic communications \cite{Telecommunication Band}, causing significant QI loss over long-distance transmission. Quantum frequency conversion (QFC) serves as a solution to manipulate the optical frequency of photons while preserving QI with high fidelity \cite{QFC}. Implementing a telecom-band QFC scheme becomes crucial to enable devices operating at non-communication frequencies to exchange QI through optical fibers with minimal loss and maximum fidelity \cite{Time-bin Encoded Qubits Quantum Nodes and Channels, Quantum Information Preserving-1, Quantum Information Preserving-2, Telecom QFC progress, Telecom QFC NV Centers, Telecom QFC Quantum Network Link Single Atoms, Telecom QFC Quantum Network Link Atomic Ensembles}.

QFC undergoes validation across various platforms and is typically implemented through one of three approaches: utilizing a $\chi^{(2)}$ nonlinear crystal, a $\chi^{(3)}$ nonlinear crystal, or a $\chi^{(3)}$ atomic ensemble. In the first approach, a $\chi^{(2)}$ nonlinear crystal is employed, with the system typically operating far from resonance, resulting in negligible spontaneous emission loss. However, this method necessitates a strong pumping laser to ensure sufficient atom-field interactions, leading to the occurrence of noise pollution \cite{Three Wave Mixing SPDC Noise Pollution, Three Wave Mixing Noise Pollution}, thereby diminishing the fidelity of the converted field. While weak pumping power requirements can be met using a cavity \cite{QFC three-wave mixing cavity-1, QFC three-wave mixing cavity-2, QFC three-wave mixing cavity-3, QFC three-wave mixing cavity-4, QFC three-wave mixing cavity-5} or a waveguide \cite{Telecom QFC three-wave mixing waveguide-1, Telecom QFC three-wave mixing waveguide-2, Telecom QFC three-wave mixing waveguide-3, Telecom QFC three-wave mixing waveguide-4, Telecom QFC three-wave mixing waveguide-5}, effectively suppressing noise pollution remains a challenging task. In the second method of employing a $\chi^{(3)}$ nonlinear crystal, the system remains far from resonance. However, to suppress the influence of the undesired symmetric-conversion channel \cite{Four Wave Mixing Symmetric Channel}, the symmetry between the channels must be broken. Feasible solutions, such as crystal fibers \cite{QFC four-wave mixing fiber-1, QFC four-wave mixing fiber-2, QFC four-wave mixing fiber-3} or microresonators \cite{QFC four-wave mixing microresonator-1, QFC four-wave mixing microresonator-2, QFC four-wave mixing microresonator-3}, introduce additional insertion photon loss, consequently reducing the overall efficiency of the conversion system. The third approach of utilizing a $\chi^{(3)}$ atomic ensemble involves lower pumping power requirements compared to other platforms, as the system operates under a near-resonant condition. This condition ensures negligible pump-induced noise pollution \cite{Similar Approach on Double-lambda QFC}. Although the atom-field interactions within the undesired conversion channel are negligibly weak, allowing for its exclusion from consideration, the near-resonant condition also results in spontaneous emission loss, consequently reducing the conversion efficiency (CE). The introduction of electromagnetically induced transparency (EIT) \cite{EIT Theory, EIT4, EIT5, QM3, HOM} efficiently suppresses disturbances from the vacuum field reservoir, leading to a reduction in spontaneous emission loss \cite{FWM1, FWM2, FWM3, FWM4}. Furthermore, EIT significantly enhances the nonlinear interaction between atoms and photons, enabling the system to achieve efficient QFC \cite{Double-lambda QFC Experiment, Energy Level and CG Coefficient and Double-lambda QFC Theorem}.

In this paper, we explore a near-resonant QFC scheme utilizing diamond-type four-wave mixing (FWM) in rubidium atomic ensembles \cite{Telecom QFC four-wave mixing atomic ensemble-1, Telecom QFC four-wave mixing atomic ensemble-2, Diamond QFC Experiment}. The suppression of spontaneous emission loss is achieved through the cascade-type EIT structure \cite{Cascade EIT} within the diamond-type system, resulting in a highly efficient conversion process. The conversion wavelengths of photons are determined by the FWM process. Specifically, we choose transition schemes that convert the wavelength of photons between the near-infrared of 795 nm and the telecom band of 1367 nm (telecom E-band) or 1529 nm (telecom C-band). The former is ideal for quantum computation \cite{Neutral Atom Quantum Computing, Neutral Atom Quantum Simulation-1, Neutral Atom Quantum Simulation-2, Neutral Atom Quantum Algorithms, Neutral Atom High-Fidelity Entangling Gates} and QI storage \cite{Light Storage Minute Scale, Quantum Memory sub-Second Scale, EIT Quantum Memory, Polarization Encoded Qubits Single Photon Quantum Memory, Vapor Cell Quantum Memory Single-Photon Storage} through rubidium atoms, while the latter is well-suited for a fiber-based QI network. Notably, this QFC scheme can be integrated with the Duan-Lukin-Cirac-Zoller (DLCZ) quantum repeater protocol \cite{DLCZ Protocol, Phase-Stabled DLCZ Protocol, DLCZ Protocol Realization}. Although a theoretical model for this QFC scheme was previously established for semi-classical quantities, such as transmittance and CE \cite{Diamond System, Diamond System Thesis}, the applied field absorption loss was neglected in that model. Our study reveals the limitation of this simplification in the high CE regime. Importantly, the previous model lacked a theoretical framework to characterize the quantum properties of this QFC system.

We utilize the Heisenberg-Langevin approach \cite{Heisenberg-Langevin method} and the reduced-density-operator method \cite{Reduce Density Matrix} to construct a quantum model that provides a quantum mechanical description of the conversion process. The general form of the ladder operators for the quantized fields is derived with consideration of both the Langevin noise and the applied field absorption loss. This derivation offers a more comprehensive description of the diamond-type QFC scheme. The transmittance and CE are derived for frequency down- and up-conversion, and we identify the parameters that maximize the CE at different optical depths (ODs). The transition schemes for optimizing the CE are selected to ensure that the transition of the $D_2$ line is a cycling transition, and the optimized CE is the maximum among all possible energy level configurations. We also argue for the necessity of considering the applied field absorption loss by analyzing the optimization curves.

Quantum properties of the converted field, including quadrature variances, photon statistics, and fidelity, are thoroughly discussed, and their exact forms are derived for any arbitrary input field. In the context of QFC, which serves as a quantum interface bridging diverse photonic wavelengths, it is crucial to highly preserve the encoded QI in qubits during the conversion process. We are the first to theoretically demonstrate that, in a diamond-type QFC system, the conversion scheme effectively retains the QI encoded in the photon-number, path, and polarization degrees of freedom (DOFs). These encoded qubits demonstrate notable entanglement retention under sufficiently high CE. In the scenario of perfect CE, the scheme achieves unity fidelity. Moreover, both the CE and fidelity remain resilient against noise introduced by vacuum fluctuations, allowing the system to implement high-purity QFC.

The paper is structured as follows. In Sec. \ref{sec:ladder operators}, we use the Heisenberg-Langevin approach to obtain the general form of the field operators for the transmitted and converted optical fields. In Sec. \ref{sec:transmittance and conversion efficiency}, we derive the CE and transmittance of the frequency down- and up-conversion, optimizing the parameters to achieve the maximum CE at different ODs. In Sec. \ref{sec:quadrature variance}, we discuss the quadrature variance of the output fields for any arbitrary input state. The quadrature variance is calculated for the n-photon Fock state, coherent state, and squeezed coherent state. In Sec. \ref{sec:converted signal state}, we use the reduced-density-operator method to obtain the quantum state of the converted field for any arbitrary input state. The density operator and the conversion fidelity are then further analyzed for the Fock and coherent input states. In Sec. \ref{sec:quantum information interface}, we discuss the retention of the QI carried by the single-rail, path, and polarization photonic qubits after the diamond-type QFC process. In Sec. \ref{sec:entanglement retention}, we present the retention of entangled qubits by extending the system to implement an N-qubit QFC. Finally, Section \ref{sec:conclusion} summarizes our findings and outlines prospects for future work. The technical details and supplementary information are provided in the appendices.


\section{Quantum Model} \label{sec:ladder operators}

\subsection{Heisenberg-Langevin Approach} \label{subsec:HLE}

We consider a cold atomic ensemble comprising diamond-type four-level atoms with a metastable ground state and three excited states, as depicted in Fig. \ref{fig:energy level diagram}(a). The strong driving and coupling fields are treated classically, and the field-dipole coupling strength is described by the Rabi frequency $\Omega_{d(c)}=2d_{42(31)}E_{d(c)}/\hbar$, where $d_{ij}$ represents the electric dipole matrix element. The weak probe and signal fields are quantized and can be described by ladder operators $\hat{a}_{p(s)}$. The four participating light fields can be multimode fields, and they are assumed to propagate in the same direction, as depicted in Fig. \ref{fig:energy level diagram}(b). All detunings between the fields and the atomic resonance, denoted as $\Delta_p$, $\Delta_c$, and $\delta$, are taken into account in the theoretical model. By using the collective atomic operator approach \cite{Collective Atomic Operator}, the system can be described collectively. Under the rotating wave approximation and the slowly varying amplitude (SVA) \cite{SVA}, the Hamiltonian of the entire system is expressed as follows:
\begin{align}
\hat{H}_S&=-\frac{N\hbar}{2L}\int_0^L \big[ 2g_p\hat{a}_p(z,t)\hat{\sigma}_{21}(z,t)+\Delta_p\hat{\sigma}_{22}(z,t)\notag\\
+&2g_s\hat{a}_s(z,t) e^{-i\Delta kz}\hat{\sigma}_{43}(z,t)+\delta\hat{\sigma}_{44}(z,t)+\Delta_c\hat{\sigma}_{33}(z,t)\notag\\
+&\Omega_c(z,t)\hat{\sigma}_{31}(z,t)+\Omega_d(z,t)\hat{\sigma}_{42}(z,t)+h.c.\big] dz. \label{eq:(1)}
\end{align}
Here, the cold atomic ensemble system we study exhibits characteristics of a low-density dilute gas, leading us to disregard the interactions between atoms. $N$ represents the total number of atoms, and $L$ denotes the length of the atomic ensemble. The phase-mismatch parameter is defined as $\Delta k=k_p+k_d-k_s-k_c$, and it is eliminated for the co-propagation case. The coupling constants between the quantized fields and electric dipoles are denoted by $g_{p(s)}=d_{21(43)}\epsilon_{p(s)}/\hbar$, where $\epsilon_{p(s)}=\sqrt{\frac{\hbar \omega_{p(s)}}{2 \epsilon_0 V}}$ represents the amplitude of the quantized fields. $\hat{\sigma}_{ij}(z,t)$ denotes the collective atomic operator after the SVA and can be obtained by solving the following Heisenberg-Langevin equations (HLEs) \cite{Quantum Optics}:
\begin{align}
\frac{\partial\hat{\sigma}_{ij}}{\partial t}=\frac{i}{\hbar}\big[\hat{H}_S,\hat{\sigma}_{ij}\big]+\hat{R}_{ij}+\hat{F}_{ij}, \label{eq:(2)}
\end{align}
where $\hat{R}_{ij}$ represents the relaxation term, and $\hat{F}_{ij}$ denotes the Langevin noise operator. Additional information on the HLEs, relaxation terms, and Langevin noise operators can be found in Appendices \ref{sec:Appendix A} and \ref{sec:Appendix B}. Due to the selection rules \cite{Selection Rules}, the transitions between energy levels $\ket{2}$ and $\ket{3}$ and between energy levels $\ket{1}$ and $\ket{4}$ do not need to be considered in the diamond-type configuration.

\FigOne

For the diamond-type system, the weak probe and signal fields can both be treated as perturbation fields. We solve the zeroth-order HLEs under the steady-state assumption for zeroth-order populations. The HLEs for $\hat{\sigma}_{11}$, $\hat{\sigma}_{13}$, $\hat{\sigma}_{22}$, $\hat{\sigma}_{24}$, $\hat{\sigma}_{31}$, $\hat{\sigma}_{33}$, $\hat{\sigma}_{42},$ and $\hat{\sigma}_{44}$ are decoupled from the others, and the zeroth-order solutions to these decoupled HLEs take the following form:
\begin{align}
\hat{\sigma}_{ij}^{(0)}(z)=\braket{\hat{\sigma}_{ij}^{(0)}(z)}+\sum_{kl}\epsilon_{kl}(z)\hat{F}_{kl}(z). \label{eq:(3)}
\end{align}
The expectation values of the above zeroth-order collective atomic operators are listed as follows:
\begin{align}
\braket{\hat{\sigma}_{11}^{(0)}(z)}&=\frac{\Gamma_{31}(\gamma_{31}^2+4 \Delta_c^2)+\gamma_{31}{|\Omega_c|}^2}{\Gamma_{31}(\gamma_{31}^2+4 \Delta_c^2)+2 \gamma_{31}{|\Omega_c|}^2}, \label{eq:(4)}\\
\braket{\hat{\sigma}_{13}^{(0)}(z)}&=\frac{i \Gamma_{31}(\gamma_{31}+2 i \Delta_c) \Omega_c}{\Gamma_{31}(\gamma_{31}^2+4 \Delta_c^2)+2 \gamma_{31}{|\Omega_c|}^2}, \label{eq:(5)}
\end{align}
$\braket{\hat{\sigma}_{33}^{(0)}(z)}=1-\braket{\hat{\sigma}_{11}^{(0)}(z)}$, and $\braket{\hat{\sigma}_{31}^{(0)}(z)}=\braket{\hat{\sigma}_{13}^{(0)}(z)}^{*}$. The expectation values of the remaining zeroth-order atomic operators are all zero. To solve for the first-order atomic operators, we substitute the zeroth-order results into the relevant first-order HLEs as follows:
\begin{align}
\frac{\partial}{\partial t}\hat{\sigma}_{12}^{(1)}&=i\bigg[\hat{a}_p g_p \braket{\hat{\sigma}_{11}^{(0)}}+\frac{1}{2}\hat{\sigma}_{14}^{(1)}\Omega_d^*-\frac{1}{2}\hat{\sigma}_{32}^{(1)}\Omega_c\notag\\
&+\Delta_p\hat{\sigma}_{12}^{(1)}\bigg]-\frac{1}{2}\gamma_{21}\hat{\sigma}_{12}^{(1)}+\hat{F}_{12},\label{eq:(6)}\\
\frac{\partial}{\partial t}\hat{\sigma}_{14}^{(1)}&=i\bigg[\hat{a}_s g_s \braket{\hat{\sigma}_{13}^{(0)}} e^{-i \Delta k z}+\frac{1}{2}\hat{\sigma}_{12}^{(1)}\Omega_d-\frac{1}{2}\hat{\sigma}_{34}^{(1)}\Omega_c\notag\\
&+\delta \hat{\sigma}_{14}^{(1)}\bigg]-\frac{1}{2}\gamma_{41}\hat{\sigma}_{14}^{(1)}+\hat{F}_{14},\label{eq:(7)}\\
\frac{\partial}{\partial t}\hat{\sigma}_{32}^{(1)}&=i\bigg[\hat{a}_p g_p\braket{\hat{\sigma}_{31}^{(0)}}+\frac{1}{2}\hat{\sigma}_{34}^{(1)}\Omega_d^*-\frac{1}{2}\hat{\sigma}_{12}^{(1)}\Omega_c^*\notag\\
&+(\Delta_p-\Delta_c)\hat{\sigma}_{32}^{(1)}\bigg]-\frac{1}{2}\gamma_{32}\hat{\sigma}_{32}^{(1)}+\hat{F}_{32},\label{eq:(8)}\\
\frac{\partial}{\partial t}\hat{\sigma}_{34}^{(1)}&=i\bigg[\hat{a}_s g_s\braket{\hat{\sigma}_{33}^{(0)}}e^{-i \Delta k z}+\frac{1}{2}\hat{\sigma}_{32}^{(1)}\Omega_d-\frac{1}{2}\hat{\sigma}_{14}^{(1)}\Omega_c^*\notag\\
&+(\delta-\Delta_c)\hat{\sigma}_{34}^{(1)}\bigg]-\frac{1}{2}\gamma_{43}\hat{\sigma}_{34}^{(1)}+\hat{F}_{34}.\label{eq:(9)}
\end{align}
Under the assumption of weak probe and signal fields, it is important to note that the terms $\hat{a}_{p(s)}\hat{F}_{ij}$ are relatively small and have been neglected. We solve the four coupled first-order HLEs [Eqs. (\ref{eq:(6)})--(\ref{eq:(9)})] by Fourier transforming them into the frequency domain, thereby obtaining the frequency-domain first-order atomic operators.


\subsection{Field Operators} \label{subsec:MSE}

To investigate the behavior of the probe and signal fields propagating in the diamond-type QFC atomic medium, we solve the following Maxwell-Schr\"{o}dinger equations (MSEs):
\begin{align}
\left(\frac{1}{c}\frac{\partial}{\partial t}+\frac{\partial}{\partial z}\right)\hat{a}_p(z,t)&=i\frac{N}{c}g_p^*\hat{\sigma}_{12}^{(1)}(z,t),\label{eq:(10)}\\
\left(\frac{1}{c}\frac{\partial}{\partial t}+\frac{\partial}{\partial z}\right)\hat{a}_s(z,t)&=i\frac{N}{c}g_s^*\hat{\sigma}_{34}^{(1)}(z,t)e^{i \Delta k z},\label{eq:(11)}\\
\left(\frac{1}{c}\frac{\partial}{\partial t}+\frac{\partial}{\partial z}\right)\Omega_c(z,t)&=\frac{i \alpha_c \Gamma_{31}}{2 L}\braket{\hat{\sigma}_{13}^{(0)}(z)},\label{eq:(12)}
\end{align}
where $\Delta k=0$ under the condition of co-propagation considered here. In the coupling field MSE [Eq. (\ref{eq:(12)})], $\alpha_{c}$ represents the OD of the coupling field, defined as $\alpha_{c}=n\sigma_{c}L$, where $n$ is the atomic density, and $\sigma_{c}$ is the scattering cross section of the coupling field. The absorption loss of the driving field is neglected because, under the condition of a weak probe field, the atomic operator involved in the driving field MSE, $\hat{\sigma}_{24}(z,t)$, is negligible. The steady-state solution to the coupling field MSE can be found in Appendix \ref{sec:Appendix C}; for simplicity, we denote it as $\Omega_c(z)$ and treat it as a function of $z$ in the subsequent derivation for the probe and signal fields. Next, by applying the Fourier transform $\hat{a}_{p(s)}(z,t)=\int_{-\infty}^{\infty} d\omega \widetilde{a}_{p(s)}(z,\omega)e^{-i\omega t}$ to the coupled MSEs [Eqs. (\ref{eq:(10)}) and (\ref{eq:(11)})] and substituting the first-order atomic operators derived from the HLEs, the MSEs for the probe and signal fields in the frequency domain can be rearranged into the following form:
\begin{align}
\frac{\partial}{\partial z}\widetilde{a}_p(z,\omega)&=\Lambda_p(z,\omega)\widetilde{a}_p(z,\omega)+\kappa_p(z,\omega)\widetilde{a}_s(z,\omega)\notag\\
&+\sum_{\alpha_i}\xi_{\alpha_i}^{p}(z,\omega)\widetilde{f}_{\alpha_i}(z,\omega),\label{eq:(13)}\\
\frac{\partial}{\partial z}\widetilde{a}_s(z,\omega)&=\Lambda_s(z,\omega)\widetilde{a}_s(z,\omega)+\kappa_s(z,\omega)\widetilde{a}_p(z,\omega)\notag\\
&+\sum_{\alpha_i}\xi_{\alpha_i}^{s}(z,\omega)\widetilde{f}_{\alpha_i}(z,\omega),\label{eq:(14)}
\end{align}
where $\Lambda_{p(s)}(z,\omega)$ denotes the self-coupling coefficient and $\kappa_{p(s)}(z,\omega)$ denotes the cross-coupling coefficient. In addition, $\widetilde{f}_{\alpha_i}(z,\omega)=\sqrt{\frac{N}{c}}\widetilde{F}_{\alpha_i}(z,\omega)$ is the normalized Langevin noise operator, which obeys the delta correlation described in Appendix \ref{sec:Appendix B}, and $\alpha_i$ denotes the $\{12,14,32,34\}$ subspace of the atomic operators \cite{Heisenberg-Langevin method}; $\xi_{\alpha_i}^{p(s)}(z,\omega)$ denotes the coefficient of noise disturbance for the system. The explicit form of the coefficients for the two coupled equations (\ref{eq:(13)}) and (\ref{eq:(14)}) can be found in Appendix \ref{sec:Appendix D}. We reformulate the coupled equations in matrix form and solve the first-order linear ordinary differential equation (ODE) for the two-dimensional linear operator with the initial condition at $z=0$. The solution to the ODE cannot be obtained directly with the integrating factor because of the $z$ dependence of the coefficient matrix $M(z,\omega)$, which arises from the loss in the coupling field $\Omega_c(z)$. The expression for the matrix $M(z,\omega)$ can be found in Appendix \ref{sec:Appendix E}. Instead, we used the approach introduced by Magnus for solving the ODE \cite{Magnus Expansion, Magnus Expansion Applications}. The general solution for the ladder operators can be obtained as follows:

\begin{align}
\begin{bmatrix}
\widetilde{a}_p(L,\omega)\\
\widetilde{a}_s(L,\omega)
\end{bmatrix}&=e^{\Omega(L,\omega)}
\begin{bmatrix}
\widetilde{a}_p(0,\omega)\\
\widetilde{a}_s(0,\omega)
\end{bmatrix}\notag\\
+e^{\Omega(L,\omega)}&\sum_{\alpha_i}\int_{0}^{L} e^{-\Omega(z,\omega)}
\begin{bmatrix}
\xi_{\alpha_i}^{p}(z,\omega)\\
\xi_{\alpha_i}^{s}(z,\omega)
\end{bmatrix}
\widetilde{f}_{\alpha_i}(z,\omega) dz, \label{eq:(15)}
\end{align}
where the parametric evolution of the diamond-type system is characterized by the term $e^{\Omega(L,\omega)}$. Its matrix form can be written as
\begin{align}
e^{\Omega(L,\omega)}&={\rm exp}\left[\sum_{n=1}^{\infty} \Omega_n(L,\omega)\right]\equiv 
\begin{bmatrix}
A(\omega) & B(\omega)\\
C(\omega) & D(\omega)
\end{bmatrix}, \label{eq:(16)}
\end{align}
where $\Omega_n(L,\omega)$ is the $n$th-order term of Magnus expansion for the two-dimensional linear ODE. The specifics of this method are provided in Appendix \ref{sec:Appendix E}. The simplified form for the field operators is as follows:
\begin{align}
\begin{bmatrix}
\widetilde{a}_p(L,\omega)\\
\widetilde{a}_s(L,\omega)
\end{bmatrix}=&\begin{bmatrix}
A(\omega) & B(\omega)\\
C(\omega) & D(\omega)
\end{bmatrix}\begin{bmatrix}
\widetilde{a}_p(0,\omega)\\
\widetilde{a}_s(0,\omega)
\end{bmatrix}\notag\\
+\sum_{\alpha_i}&\int_{0}^{L} \begin{bmatrix}
P_{\alpha_i}(z,\omega)\\
Q_{\alpha_i}(z,\omega)
\end{bmatrix}\widetilde{f}_{\alpha_i}(z,\omega) dz, \label{eq:(17)}
\end{align}
where $P(Q)_{\alpha_i}(z,\omega)$ characterizes the influence of noise on the evolution of the system. Having derived the ladder operators for the probe and signal fields, we can proceed to discuss the CE and transmittance for both frequency down- and up-conversion cases. The quantum properties of the diamond-type QFC system, including quadrature variances, converted photon statistics, and squeezed states, will be analyzed in the subsequent sections.


\section{Conversion Efficiency} \label{sec:transmittance and conversion efficiency}

\subsection{Steady-State Condition} \label{subsec:Steady-State Condition}

For simplicity, we assume that the input field has reached a steady state; therefore, here we focus solely on the single-frequency-mode behavior of the QFC system. The ladder operators still follow the relation as in Eq. (\ref{eq:(17)}), whereas the operators are replaced by $\widetilde{a}_{p(s),\omega}(z)$ and $\widetilde{f}_{\alpha_i,\omega}(z)$. These discretized operators are equipped with both the commutation relation and the delta correlation \cite{Quantum Optics} as follows:
\begin{align}
&\big[\widetilde{a}_{p(s),\omega}(z),\widetilde{a}_{p(s),\omega'}^{\dagger}(z)\big]=\delta_{\omega\omega'}, \label{eq:(18)}\\
&\braket{\widetilde{f}_{\alpha_i,\omega}(z)\widetilde{f}_{\alpha_j,\omega'}(z')}=\delta_{\omega\omega'}D_{\alpha_i,\alpha_j}\delta(z-z'). \label{eq:(19)}
\end{align}
Firstly, our attention is focused on the frequency down-conversion from the probe field to the signal field. The signal field is initially in the vacuum state $\ket{0}$, and then at $z=L$, the photon number of the signal field becomes:
\begin{align}
n_{s,\omega}(L)=\braket{\widetilde{a}_{s,\omega}^{\dagger}(L)\widetilde{a}_{s,\omega}(L)}. \label{eq:(20)}
\end{align}
By substituting the single-mode version of Eq. (\ref{eq:(17)}) into Eq. (\ref{eq:(20)}) and applying the delta correlation of the Langevin noise operators [Eq. (\ref{eq:(19)})], the expression for the signal photon number can be derived as follows:
\begin{align}
n_{s,\omega}(L)=&|C(\omega)|^2\braket{\widetilde{a}_{p,\omega}^{\dagger}(0)\widetilde{a}_{p,\omega}(0)}\notag\\
+&\sum_{\alpha_i,\alpha_j}\int_0^L dz Q_{\alpha_i}^*(z,\omega)Q_{\alpha_j}(z,\omega)D_{\alpha_i^{\dagger},\alpha_j}. \label{eq:(21)}
\end{align}
Here, $\alpha_i^{\dagger}$ denotes the $\{21,41,23,43\}$ subspace of the adjoint atomic operators and $\widetilde{f}_{\alpha_i,\omega}^{\dagger}(z)=\widetilde{f}_{\alpha_i^{\dagger},\omega}(z)$. In a similar manner, the probe photon number at $z=L$ can be obtained as follows:
\begin{align}
n_{p,\omega}(L)=&|A(\omega)|^2\braket{\widetilde{a}_{p,\omega}^{\dagger}(0)\widetilde{a}_{p,\omega}(0)}\notag\\
+&\sum_{\alpha_i,\alpha_j}\int_0^L dz P_{\alpha_i}^*(z,\omega)P_{\alpha_j}(z,\omega)D_{\alpha_i^{\dagger},\alpha_j}. \label{eq:(22)}
\end{align}
The normal-order diffusion coefficients $D_{\alpha_i^{\dagger},\alpha_j}$ can be derived using the Einstein relation \cite{Quantum Optics} in conjunction with the zeroth- and first-order atomic operators. By dropping all of the higher-order ($\geq2$) perturbation terms, all diffusion coefficients are calculated to be zero. Further details regarding the Einstein relations and the diffusion coefficients are provided in Appendix \ref{sec:Appendix B}. In the steady-state condition, we can set the frequency $\omega$ to zero, as the part where $\omega$ equals zero covers all the results from the discretized frequency distribution. Finally, the transmittance of the probe field and the CE of the signal field for frequency down-conversion can be obtained as follows:
\begin{align}
&T_{d}=\frac{n_{p,0}(L)}{n_{p,0}(0)}=|A(0)|^2, \label{eq:(23)}\\
&\eta_{d}=\frac{n_{s,0}(L)}{n_{p,0}(0)}=|C(0)|^2. \label{eq:(24)}
\end{align}
For the up-conversion case, where the signal field is converted to the probe field and the probe field is initially in the vacuum state $\ket{0}$, a similar derivation yields the following transmittance for the signal field and CE for the probe field:
\begin{align}
&T_{u}=\frac{n_{s,0}(L)}{n_{s,0}(0)}=|D(0)|^2, \label{eq:(25)}\\
&\eta_{u}=\frac{n_{p,0}(L)}{n_{s,0}(0)}=|B(0)|^2. \label{eq:(26)}
\end{align}


\subsection{Conversion Efficiency Maximization} \label{subsec:conversion efficiency optimization}

We plot the optimized CE curve for frequency down-conversion using a $\prescript{87}{}{\rm Rb}$ atomic ensemble, as shown in Fig. \ref{fig:Optimized Curve}. The CE optimization for each OD is achieved by adjusting five parameters: $\Delta_p$, $\Delta_c$, $\delta$, $\Omega_c$, and $\Omega_d$. This adjustment is carried out in a way that maximizes the CE while ensuring a continuous variation of the optimized parameters as the OD increases. We can always find a higher CE for any OD by continuously increasing the detunings and Rabi frequencies while adjusting them appropriately; however, the resulting increase in CE is minuscule. On the other hand, by constraining the parameter ranges, we can also identify other suitable combinations of parameters that achieve only a slightly lower CE. 

The energy level configurations and transition schemes of the telecom E-band and C-band QFC are carefully selected to ensure that the transition between $\ket{1}$ and $\ket{3}$ ($D_2$ line, 780 nm) constitutes a cycling transition and that the CE is maximized among all possible energy level choices. A cycling transition is crucial to prevent the population of the excited sub-Zeeman state from decaying to the sub-Zeeman states outside its corresponding transition, commonly referred to as the dark states. For the telecom E-band QFC scheme \cite{Rubidium 87 D Line Data and CG Coefficients, Rubidium Energy Level Data}, operating at wavelengths between 795 and 1367 nm, we have selected the following energy levels: $\ket{1}=\ket{5S_{1/2}, F=2, m_F=2}$, $\ket{2}=\ket{5P_{1/2}, F=1, m_F=1}$, $\ket{3}=\ket{5P_{3/2}, F=3, m_F=3}$, and $\ket{4}=\ket{6S_{1/2}, F=2, m_F=2}$. Similarly, for the telecom C-band QFC scheme \cite{Rubidium 87 D Line Data and CG Coefficients, Rubidium Energy Level Data, 4D3/2 Hyperfine Splitting and CG Coefficient, 5P3/2-4D3/2 Transition Energy}, operating between 795 and 1529 nm, the selected energy levels are: $\ket{1}=\ket{5S_{1/2}, F=2, m_F=2}$, $\ket{2}=\ket{5P_{1/2}, F=1, m_F=1}$, $\ket{3}=\ket{5P_{3/2}, F=3, m_F=3}$, and $\ket{4}=\ket{4D_{3/2}, F=2, m_F=2}$. It is important to note that the transition schemes and optimization results using a $\prescript{85}{}{\rm Rb}$ atomic ensemble are similar to those obtained using $\prescript{87}{}{\rm Rb}$.


\FigTwo

\TableOne

Since there exists only one possible transition scheme for sub-Zeeman level transition, we can use the results in Sec. \ref{subsec:Steady-State Condition} to maximize the CE. The spontaneous decay rate of the fine-structure transition $\Gamma_{\lambda_{J', J}}$ ($\lambda_{J', J}$ is the transition wavelength) is related to the transition rate between hyperfine sub-Zeeman states $\Gamma_{ij}$ by square of the transition coefficient (similar to the Clebsch-Gordan coefficient from the spin-orbital coupling), denoted as $a_{J',F',m_F'\rightarrow J,F,m_F}$ \cite{Transfer Coefficient, CG Coefficient Table 1}. The transition coefficient follows the relation $\braket{F',m_F'|e\vec{r}|F,m_F}=a_{J',F',m_F'\rightarrow J,F,m_F}\braket{J'||e\vec{r}||J}\hat{e}_{m_F'-m_F}$ \cite{Rubidium 87 D Line Data and CG Coefficients, Theorem about Spontaneous Decay, Wigner-Eckart Theorem, 3-j and 6-j Symbol}, where the reduced matrix element $\braket{J'||e\vec{r}||J}$ \cite{Theorem about Spontaneous Decay, Experimental Reduced Matrix Element} obeys the asymmetric convention. The OD in this section, denoted by $\alpha$, is defined as $\alpha=n\sigma_p L |a_{1/2,1,1\rightarrow 1/2,2,2}|^2$, where $\sigma_p$ is the scattering cross section of the probe field; note that the OD satisfies $\alpha=\frac{4LN|g_p|^2}{c\Gamma_{780}}$, where $\Gamma_{780}=2\pi\times 6.063$ MHz. This OD can be determined through experiments.

The optimized CE curve for the telecom E-band QFC scheme is depicted in Fig. \ref{fig:Optimized Curve}(a). When $\alpha=240$, the CE reaches $90\%$ and asymptotically approaches $100\%$ with increasing OD. In Fig. \ref{fig:Optimized Curve}(b), we present the optimized CE curve for the telecom C-band QFC scheme, reaching $80\%$ at $\alpha=700$. It is noteworthy that, under the same OD, the E-band scheme exhibits a higher optimized CE compared to the C-band scheme. This is primarily due to the differences in the spontaneous emission rate $\Gamma_{43}=|a_{J',F',m_F'\rightarrow J,F,m_F}|^2\Gamma_{\lambda_{J', J}}$, determined by the fine-structure transition rates and the transition coefficients between $\ket{3}$ and $\ket{4}$ for these two schemes. A higher spontaneous emission rate generally indicates stronger dipole-field interactions, thereby leading to a greater CE.

Table \ref{table:optimized parameters} displays the optimized parameter sets that maximize CEs for various ODs. Although the requirements for detunings and Rabi frequencies may appear stringent, constraining the parameter scanning ranges allows us to identify alternative parameter sets with slightly lower CE but more easily achievable conditions. Achieving a large OD is crucial for highly efficient QFC, and the OD conditions specified in Table \ref{table:optimized parameters} are experimentally attainable \cite{Large OD-1, Large OD-2, Large OD-3}.

The blue and dotted curves in Fig. \ref{fig:Optimized Curve} indicate the necessity of considering coupling field absorption loss in the diamond-type QFC system. While this effect has minimal impact on the model in the low CE (or OD) regime \cite{Diamond System}, its significance becomes pronounced in the high CE regime. For instance, at $\alpha=700$ for the telecom C-band QFC scheme, the difference amounts to $3.5\%$, emphasizing its critical role in practical applications. Utilizing a nonabsorbing model for calculating optimized parameters would yield misleading results, not aligning with the optimal CE in real-world systems. Our findings underscore the superior accuracy of our model compared to other nonabsorbing models in predicting CE in the high CE regime. The heightened precision in predictions holds considerable implications for practical applications, such as reducing entanglement distribution time in certain quantum repeater protocols \cite{Quantum Repeaters}.

Here, we provide a physical picture for the optimized results, which shares similarities with some earlier proposed models \cite{Telecom QFC four-wave mixing atomic ensemble-1, Diamond System}. To maximize CE, establishing a strong correlation between $\ket{3}$ and $\ket{4}$ is crucial. For this purpose, we must ensure that both the cascade-type EIT (composed of $\ket{1}$, $\ket{2}$, and $\ket{4}$) and the correlation between $\ket{1}$ and $\ket{3}$ (established by the coupling field) are sufficiently robust. Achieving a powerful cascade-type EIT can significantly reduce the spontaneous emission loss of the probe field, but it requires a nearly resonant driving field ($\Delta_d=\delta-\Delta_p\approx 0$) with a large $\Omega_d$. Additionally, to establish a strong correlation between $\ket{1}$ and $\ket{3}$ while simultaneously suppressing spontaneous decay, the coupling field needs to have large values of $\Omega_c$ and $\Delta_c$. However, this induces an AC Stark shift on $\ket{1}$ and $\ket{3}$, necessitating the introduction of a corresponding probe detuning ($\Delta_p$) to maintain the two-photon resonance in cascade-type EIT, protecting the probe field and maximizing CE. For larger OD conditions, larger $\Omega_d$ is necessary to reduce the spontaneous emission loss of the probe field, resulting in increased optimal values for the five parameters as OD increases.

The aforementioned physical picture aligns with the optimized parameters identified in our simulation (Table \ref{table:optimized parameters}). This alignment is confirmed by examining the AC Stark shift of the corresponding energy levels. Notably, each optimized CE curve has two sets of optimized parameters, determined by the direction of the AC Stark shift. While the magnitudes of the five parameters remain the same, the detunings have opposite signs. Interestingly, for frequency up-conversion, the optimized five parameters and the corresponding CEs mirror those of the down-conversion case. This observation implies that the same set of parameters can maximize the CE for both frequency down- and up-conversion in the diamond-type QFC system.


\section{Quadrature Variance} \label{sec:quadrature variance}

\subsection{General Formula} \label{subsec:General Form}

In this section, we derive the general formulas for the quadrature variances of the transmitted and converted fields in both the down- and up-conversion cases. The two quadrature operators are defined as follows:
\begin{align}
&\widetilde{X}_{p(s)}(z)=\frac{1}{2}\big[\widetilde{a}_{p(s)}(z)+\widetilde{a}_{p(s)}^{\dagger}(z)\big], \label{eq:(27)}\\
&\widetilde{Y}_{p(s)}(z)=\frac{1}{2i}\big[\widetilde{a}_{p(s)}(z)-\widetilde{a}_{p(s)}^{\dagger}(z)\big], \label{eq:(28)}
\end{align}
where we omit the symbol $\omega$ because, in the steady-state condition, all $\omega$ can be replaced by 0. First, we consider the down-conversion case. Following the method introduced in Appendix \ref{sec:Appendix B}, we deduce that all diffusion coefficients, $D_{\alpha_i,\alpha_j}$, $D_{\alpha_i^{\dagger},\alpha_j^{\dagger}}$, and $D_{\alpha_i^{\dagger},\alpha_j}$, are zero. By using a similar approach as in Sec. \ref{sec:transmittance and conversion efficiency}, we obtain all expectation values of one ladder operator and the multiplications of ladder operators at position $z=L$ for the probe and signal fields. The quadrature variances, following the definition in \cite{Introductory Quantum Optics}, can be derived using the previously obtained expectation values of ladder operators and can be simplified into the following forms:
\begin{align}
{\rm var}[X_{p}(L)]=&\frac{1}{4}\left\{\big\langle\big[A(0)\widetilde{a}_{p}(0)+A^*(0)\widetilde{a}_{p}^{\dagger}(0)\big]^2\big\rangle\right.\notag\\
&\left.-\big[\big\langle A(0)\widetilde{a}_{p}(0)+A^*(0)\widetilde{a}_{p}^{\dagger}(0)\big\rangle\big]^2\right\}\notag\\
&+\frac{1}{4}[1-|A(0)|^2], \label{eq:(29)}\\
{\rm var}[Y_{p}(L)]=&-\frac{1}{4}\left\{\big\langle\big[A(0)\widetilde{a}_{p}(0)-A^*(0)\widetilde{a}_{p}^{\dagger}(0)\big]^2\big\rangle\right.\notag\\
&\left.-\big[\big\langle A(0)\widetilde{a}_{p}(0)-A^*(0)\widetilde{a}_{p}^{\dagger}(0)\big\rangle\big]^2\right\}\notag\\
&+\frac{1}{4}[1-|A(0)|^2], \label{eq:(30)}\\
{\rm var}[X_{s}(L)]=&\frac{1}{4}\left\{\big\langle\big[C(0)\widetilde{a}_{p}(0)+C^*(0)\widetilde{a}_{p}^{\dagger}(0)\big]^2\big\rangle\right.\notag\\
&\left.-\big[\big\langle C(0)\widetilde{a}_{p}(0)+C^*(0)\widetilde{a}_{p}^{\dagger}(0)\big\rangle\big]^2\right\}\notag\\
&+\frac{1}{4}[1-|C(0)|^2], \label{eq:(31)}
\end{align}
\begin{align}
{\rm var}[Y_{s}(L)]=&-\frac{1}{4}\left\{\big\langle\big[C(0)\widetilde{a}_{p}(0)-C^*(0)\widetilde{a}_{p}^{\dagger}(0)\big]^2\big\rangle\right.\notag\\
&\left.-\big[\big\langle C(0)\widetilde{a}_{p}(0)-C^*(0)\widetilde{a}_{p}^{\dagger}(0)\big\rangle\big]^2\right\}\notag\\
&+\frac{1}{4}[1-|C(0)|^2]. \label{eq:(32)}
\end{align}
Subsequently, we introduce the definitions $\widetilde{a}_{p, 1}(0)=A(0)\widetilde{a}_{p}(0)$ and $\widetilde{a}_{p, 2}(0)=C(0)\widetilde{a}_{p}(0)$. The expressions can then be further streamlined as follows:
\begin{align}
&{\rm var}[X_{p}(L)]={\rm var}[X_{p, 1}(0)]+\frac{1}{4}\left[1-|A(0)|^2\right], \label{eq:(33)}\\
&{\rm var}[Y_{p}(L)]={\rm var}[Y_{p, 1}(0)]+\frac{1}{4}\left[1-|A(0)|^2\right], \label{eq:(34)}\\
&{\rm var}[X_{s}(L)]={\rm var}[X_{p, 2}(0)]+\frac{1}{4}\left[1-|C(0)|^2\right], \label{eq:(35)}\\
&{\rm var}[Y_{s}(L)]={\rm var}[Y_{p, 2}(0)]+\frac{1}{4}\left[1-|C(0)|^2\right]. \label{eq:(36)}
\end{align}
Here, the quadrature variances on the right-hand side represent the variances of the newly defined fields. If we introduce a phase shifter $\widetilde{P}(\phi_p)$, the ladder operators after applying the phase shifter can be obtained by simply adding another phase to the original output ladder operators, i.e., $\widetilde{a}_{p(s)}'(L)=\widetilde{a}_{p(s)}(L)e^{-i\phi_p}$ \cite{Phase Shifter}. By introducing a phase shifter with $\phi_p=\theta$, where $\theta$ is the phase of the mode-converted coefficient $C(0)=|C(0)|e^{i\theta}$, the quadrature variances of the output signal field can be expressed as follows:
\begin{align}
&{\rm var}[X_{s}'(L)]=\eta_d\,{\rm var}[X_{p}(0)]+\frac{1}{4}(1-\eta_d), \label{eq:(37)}\\
&{\rm var}[Y_{s}'(L)]=\eta_d\,{\rm var}[Y_{p}(0)]+\frac{1}{4}(1-\eta_d). \label{eq:(38)}
\end{align}
The derivation for the up-conversion case is analogous; the result can be obtained by interchanging the ladder operators of the probe and signal in Eqs. (\ref{eq:(29)})--(\ref{eq:(36)}) and replacing $A(0)$ and $C(0)$ with $B(0)$ and $D(0)$, respectively. If a phase shifter is applied to eliminate the phase of $B(0)$, the result can be obtained by interchanging the symbols $p$ and $s$ and replacing $\eta_d$ with $\eta_u$ in Eqs. (\ref{eq:(37)}) and (\ref{eq:(38)}).


\subsection{Fock, Coherent, and Squeezed States} \label{subsec:quadrature variance examples}

Having derived the general formula for quadrature variances, let's delve into specific cases. In this context, we consider scenarios where the input probe field is in a Fock state, coherent state, or squeezed coherent state. Assuming the input probe field is in an n-photon Fock state, $\rho_p(0)=\ket{n}\bra{n}$, the quadrature variances can be obtained from Eqs. (\ref{eq:(29)})--(\ref{eq:(32)}) using the raising and lowering properties of the ladder operators:
\begin{align}
{\rm var}[X_{p}(L)]=&{\rm var}[Y_{p}(L)]=\frac{1}{4}[1-T_d+(1+2n)T_d], \label{eq:(39)}\\
{\rm var}[X_{s}(L)]=&{\rm var}[Y_{s}(L)]=\frac{1}{4}[1-\eta_d+(1+2n)\eta_d], \label{eq:(40)}
\end{align}
which are the sums of the quadrature variances of the vacuum and the n-photon Fock states; the proportion between them depends on the transmittance [Eq. (\ref{eq:(39)})] or CE [Eq. (\ref{eq:(40)})]. For an input probe field in a single-photon Fock state $\ket{1}$, the quadrature variances of the converted signal field vary with the CE, as shown in Fig. \ref{fig:quadrature variance}(a). If the input probe field is in a coherent state, i.e., $\rho_p(0) = \ket{\beta}\bra{\beta}$, then by utilizing the relation $\widetilde{a}_p(0)\ket{\beta} = \beta\ket{\beta}$, the following quadrature variances can be obtained:
\begin{align}
{\rm var}[X_{p}(L)]&={\rm var}[Y_{p}(L)]=\frac{1}{4}, \label{eq:(41)}\\
{\rm var}[X_{s}(L)]&={\rm var}[Y_{s}(L)]=\frac{1}{4}, \label{eq:(42)}
\end{align}
which are equivalent to the vacuum variance. In the following section, we will derive the quantum state of the converted field for a coherent input; the resulting state remains a coherent state.

\FigThree

If the input probe field is in a squeezed coherent state, denoted as $\rho_p(0)=\ket{\alpha,\xi}$ with $\xi=re^{i\phi}$, then, by leveraging the operational properties of the interaction between ladder operators and the squeezed coherent states \cite{Introductory Quantum Optics}, we can obtain the quadrature variances as follows:
\begin{align}
{\rm var}[X_{p}(L)]=&\frac{1}{4}\left\{1+|A(0)|^2[\cosh(2r)-1]\right.\notag\\
-&\frac{1}{2}\left.\left[(A(0))^2e^{i\phi}+(A^*(0))^2e^{-i\phi}\right]\sinh(2r)\right\}, \label{eq:(43)}\\
{\rm var}[Y_{p}(L)]=&\frac{1}{4}\left\{1+|A(0)|^2[\cosh(2r)-1]\right.\notag\\
+&\frac{1}{2}\left.\left[(A(0))^2e^{i\phi}+(A^*(0))^2e^{-i\phi}\right]\sinh(2r)\right\}, \label{eq:(44)}\\
{\rm var}[X_{s}(L)]=&\frac{1}{4}\left\{1+|C(0)|^2[\cosh(2r)-1]\right.\notag\\
-&\frac{1}{2}\left.\left[(C(0))^2e^{i\phi}+(C^*(0))^2e^{-i\phi}\right]\sinh(2r)\right\}, \label{eq:(45)}
\end{align}
\begin{align}
{\rm var}[Y_{s}(L)]=&\frac{1}{4}\left\{1+|C(0)|^2[\cosh(2r)-1]\right.\notag\\
+&\frac{1}{2}\left.\left[(C(0))^2e^{i\phi}+(C^*(0))^2e^{-i\phi}\right]\sinh(2r)\right\}. \label{eq:(46)}
\end{align}
If we introduce a phase shifter to nullify the phase of the mode-converted coefficient $C(0)$, and the input probe field is squeezed with $\phi=0$, then the quadrature variances of the converted signal field, post the phase shifter application, can be obtained as follows:
\begin{align}
{\rm var}[X_{s}'(L)]&=\frac{1}{4}\left(1-\eta_d+\eta_d e^{-2r}\right), \label{eq:(47)}\\
{\rm var}[Y_{s}'(L)]&=\frac{1}{4}\left(1-\eta_d+\eta_d e^{2r}\right). \label{eq:(48)}
\end{align}
Here, the quadrature variances transition from those of the vacuum to the input squeezed coherent state as the CE changes from zero to one. The quadrature variances of the converted signal field for an input probe with 6-dB squeezing are illustrated in Fig. \ref{fig:quadrature variance}(b). The derivation for the up-conversion case follows a similar process. The results can be obtained by replacing $A(0)$ with $B(0)$, $C(0)$ with $D(0)$, $T_d$ with $\eta_u$, and $\eta_d$ with $T_u$ in Eqs. (\ref{eq:(39)})--(\ref{eq:(46)}). If a phase shifter is applied to eliminate the phase of $B(0)$, the results can be obtained by replacing the symbol $s$ with $p$ and $\eta_d$ with $\eta_u$ in Eqs. (\ref{eq:(47)}) and (\ref{eq:(48)}). The quadrature variances of the up-converted probe field for the input signal field in a single-photon Fock state or a 6-dB squeezed coherent state are equivalent to those depicted in Fig. \ref{fig:quadrature variance} for the down-conversion case.


\section{Converted Quantum State} \label{sec:converted signal state}

\subsection{Reduced Density Operator} \label{subsec:Reduced Density Operator}

We employ the reduced-density-operator approach to derive the quantum state of the converted field for both the down- and up-conversion processes \cite{Reduce Density Matrix}. In the Schr\"{o}dinger picture, the output state of the combined system, which includes the QFC system and the reservoir, can be expressed as
\begin{align}
&\rho_f=U\rho_i U^{\dagger}, \label{eq:(49)}
\end{align}
where $\rho_i=\rho_s(0)\otimes\rho_p(0)\otimes\rho_R$ is the initial density operator. $U$ represents the evolution operator of the combined system. The evolutions of ladder operators in the Heisenberg picture are also described by the operator $U$ as follows:
\begin{align}
\widetilde{a}_{p}(L)=tr_str_R\left\{U^{\dagger}[I_s\otimes\widetilde{a}_{p}(0)\otimes I_R]U\right\}, \label{eq:(50)}\\
\widetilde{a}_{s}(L)=tr_ptr_R\left\{U^{\dagger}[\widetilde{a}_{s}(0)\otimes I_p\otimes I_R]U\right\}. \label{eq:(51)}
\end{align}
The Schr\"{o}dinger picture density operator for the output field can be expanded with respect to the number basis. For frequency down-conversion, the density matrix element of the converted signal field can be obtained as follows:
\begin{align}
\rho_{s,mn}(L)&=\prescript{}{s}{\bra{m}}\rho_s(L)\ket{n}_s\notag\\
&=\prescript{}{s}{\bra{m}}tr_p tr_R(U\rho_i U^{\dagger})\ket{n}_s\notag\\
&=tr_s\left\{\ket{n}_s\prescript{}{s}{\bra{m}}tr_p tr_R(U\rho_i U^{\dagger})\right\}\notag\\
&=tr_s\left\{tr_p tr_R\left[(\ket{n}_s\prescript{}{s}{\bra{m}}\otimes I_p\otimes I_R)U\rho_i U^{\dagger}\right]\right\}\notag\\
&=tr\left\{U^{\dagger}(\ket{n}_s\prescript{}{s}{\bra{m}}\otimes I_p\otimes I_R)U\rho_i\right\}, \label{eq:(52)}
\end{align}
where we can define $\hat{\rho}_{s,mn}(L)=U^{\dagger}(\ket{n}_s\prescript{}{s}{\bra{m}}\otimes I_p\otimes I_R)U$, which represents the Heisenberg picture operator of the density matrix element. By utilizing the following property of ladder operators \cite{Vacuum Outer Product Property}:
\begin{align}
\sum_{l=0}^{\infty}\frac{(-1)^l}{l!}(\widetilde{a}^{\dagger})^l(\widetilde{a})^l=\ket{0}\bra{0}, \label{eq:(53)}
\end{align}
we can express the outer product $\ket{n}_s\prescript{}{s}{\bra{m}}$ as the sum of multiplication of the initial ladder operators. Thus, $\hat{\rho}_{s,mn}(L)$ can be expressed as follows:
\begin{align}
&\hat{\rho}_{s,mn}(L)\notag\\
&=U^{\dagger}\left\{\sum_{l=0}^{\infty}\chi_{mnl}[\widetilde{a}_{s}^{\dagger}(0)]^{l+n}[\widetilde{a}_{s}(0)]^{l+m}\otimes I_p\otimes I_R\right\}U\notag\\
&=\sum_{l=0}^{\infty}\chi_{mnl}[\widetilde{a}_{s}^{\dagger}(L)]^{l+n}[\widetilde{a}_{s}(L)]^{l+m}, \label{eq:(54)}
\end{align}
where we make use of the unitary property of the evolution operator and introduce $\chi_{mnl}\equiv\frac{(-1)^l}{l!}\frac{1}{\sqrt{n!m!}}$. Through further derivation, we obtain the density matrix element of the converted signal field as follows:
\begin{align}
&\rho_{s,mn}(L)=\sum_{l=0}^{\infty}\chi_{mnl}\braket{[\widetilde{a}_{s}^{\dagger}(L)]^{l+n}[\widetilde{a}_{s}(L)]^{l+m}}\notag\\
&=\sum_{l=0}^{\infty}\chi_{mnl}tr_p\left\{[C^*(0)\widetilde{a}_{p}^{\dagger}(0)]^{l+n}[C(0)\widetilde{a}_{p}(0)]^{l+m}\rho_p(0)\right\}. \label{eq:(55)}
\end{align}
The detailed derivations of Eqs. (\ref{eq:(53)}) and (\ref{eq:(55)}) can be found in Appendix \ref{sec:Appendix F}. The derivation for the frequency up-conversion is similar; the result can be obtained by interchanging all of the $s$ and $p$ symbols and replacing $C(0)$ with $B(0)$ in Eqs. (\ref{eq:(52)})--(\ref{eq:(55)}).


\subsection{Fock State and Coherent State} \label{subsec:quantum state examples}

By utilizing Eq. (\ref{eq:(55)}), we can derive the exact form of the converted signal state for any arbitrary input probe field. In this context, we examine the input probe field in either a Fock state or a coherent state, determining the density operator of the converted signal field. If the input probe field is in a Fock state, with $\rho_p(0)=\ket{q}\bra{q}$, the density matrix element for the converted signal field can be obtained as follows:
\begin{align}
&\rho_{s,mn}(L)\notag\\
&=\sum_{l=0}^{\infty}\chi_{mnl}\bra{q}[C^*(0)\widetilde{a}_{p}^{\dagger}(0)]^{l+n}[C(0)\widetilde{a}_{p}(0)]^{l+m}\ket{q}\notag\\
&=\delta_{mn}C^q_n\eta_d^n(1-\eta_d)^{q-n}(1-\delta_{\eta_d, 1})+\delta_{mn}\delta_{qn}\delta_{\eta_d, 1}. \label{eq:(56)}
\end{align}
The above expression is valid for $n, m \leq q$; otherwise, $\rho_{s, mn}(L) = 0$. Here, $C^q_n = \frac{q!}{n!(q-n)!}$ represents the binomial coefficient. The fidelity, as defined in \cite{Fidelity}, between the converted signal and the input probe can be expressed as follows:
\begin{align}
F[\rho_s(L),\rho_p(0)]=\sqrt{\braket{q|\rho_s(L)|q}}=\sqrt{\eta_d}^q. \label{eq:(57)}
\end{align}
If we input a single-photon Fock state, the converted density operator can be expressed as
\begin{align}
\rho_s(L)&=(1-\eta_d)\ket{0}\bra{0}+\eta_d\ket{1}\bra{1}, \label{eq:(58)}
\end{align}
which is a mixed state comprised of the vacuum state $\ket{0}$ and the single-photon Fock state $\ket{1}$, with the probability determined by the CE of the down-conversion process. The conversion fidelity for a single-photon Fock input state is depicted in Fig. \ref{fig:fidelity}.
\FigFour

If the input probe field is in a coherent state, $\rho_p(0)=\ket{\beta}\bra{\beta}$, then the density matrix element for the converted signal field can be obtained as follows:
\begin{align}
&\rho_{s,mn}(L)\notag\\
&=\sum_{l=0}^{\infty}\chi_{mnl}\bra{\beta}[C^*(0)\widetilde{a}_{p}^{\dagger}(0)]^{l+n}[C(0)\widetilde{a}_{p}(0)]^{l+m}\ket{\beta}\notag\\
&=\sum_{l=0}^{\infty}\chi_{mnl}[C^*(0)\beta^*]^{l+n}[C(0)\beta]^{l+m}\notag\\
&=e^{-|C(0)\beta|^2}\frac{[C(0)\beta]^m[C^*(0)\beta^*]^n}{\sqrt{m!n!}}, \label{eq:(59)}
\end{align}
and the density operator can be expressed as
\begin{align}
\rho_s(L)&=\sum_{m,n}\ket{m}\bra{n}e^{-|C(0)\beta|^2}\frac{[C(0)\beta]^m[C^*(0)\beta^*]^n}{\sqrt{m!n!}}\notag\\
&=\ket{C(0)\beta}\bra{C(0)\beta}, \label{eq:(60)}
\end{align}
which remains a coherent state $\ket{C(0)\beta}$. If we introduce a phase shifter to eliminate the phase of the mode-converted coefficient $C(0)$, the fidelity between the converted signal and the input probe is
\begin{align}
F[\rho_s(L),\rho_p(0)]&=|\braket{\beta|\sqrt{\eta_d}\beta}|=e^{-\frac{1}{2}|\beta|^2(1-\sqrt{\eta_d})^2}. \label{eq:(61)}
\end{align}
The conversion fidelities for coherent input states with one and ten photons are depicted in Fig. \ref{fig:fidelity}. A coherent input state with $\beta=1$ exhibits higher fidelity compared to $\beta=10$ for the same CE; this result is reasonable since, in phase space, the coherent state with fewer photons is closer to the origin. The distance between the converted and input states in phase space for $\beta=1$ is shorter than that for $\beta=10$. The derivation for the frequency up-conversion is similar; the result can be obtained by interchanging the $s$ and $p$ symbols and replacing $C(0)$ with $B(0)$ and $\eta_d$ with $\eta_u$ in Eqs. (\ref{eq:(56)})--(\ref{eq:(61)}). The conversion fidelities for the up-conversion cases are the same as those depicted in Fig. \ref{fig:fidelity} for the down-conversion cases.


\section{Qubit Retention} \label{sec:quantum information interface}

\subsection{Single-Rail-Encoded Qubit} \label{subsec:photon number encoded qubit}

Consider a qubit that is physically implemented by a spatial mode of the electromagnetic field (serving as the QI carrier), where the two-dimensional Hilbert space is spanned by the vacuum and one-photon Fock state \cite{Photon Number Qubit Teleportation-1,Photon Number Qubit Teleportation-2}. Quantum entanglement is shared between different modes of the electromagnetic field. The logical basis $\ket{0}$ and $\ket{1}$ are defined as the vacuum state and the one-photon Fock state, respectively; such a qubit is called the single-rail qubit \cite{Single-Rail Qubit Logic Gate, Single-Rail Qubit Passive Quantum Computing}, a particular case of photon-number encoding. In many promising QI processing systems, such as quantum dots, superconducting circuits, and single-atoms, the qubits are naturally converted into the single-rail qubits when the systems are coupled to light \cite{QIPC Europe, Single-Rail and Dual-Rail Interface}. Through the diamond-type QFC scheme, we can convert the carrier's frequency while preserving the encoded QI. Consider a frequency down-conversion from the probe to the signal, where we input the probe field with an arbitrary one-qubit state using the single-rail encoding. The density operator of the input probe field using the logical basis representation is as follows:
\begin{align}
\rho_p(0)=
\begin{bmatrix}
\rho_{00} & \rho_{01}\\
\rho_{10} & \rho_{11}
\end{bmatrix}. \label{eq:(62)}
\end{align}
The parameter $\rho_{ij}$ represents the density matrix element corresponding to $\ket{i}\bra{j}$ basis. By utilizing Eq. (\ref{eq:(55)}), we can obtain the density matrix element of the converted signal state, and its exact form is as follows:
\begin{align}
\rho&_{s,mn}(L)\notag\\
=&\sum_{l=0}^{\infty} \chi_{mnl}\sum_{i,j=0}^1 \bra{i}[C^*(0)\widetilde{a}_{p}^{\dagger}(0)]^{l+n}[C(0)\widetilde{a}_{p}(0)]^{l+m}\rho_{ji}\ket{j}\notag\\
=&\delta_{m0}\delta_{n0}\left[\rho_{00}+(1-|C(0)|^2)\rho_{11}\right]+\delta_{m0}\delta_{n1}C^*(0)\rho_{01}\notag\\
&+\delta_{m1}\delta_{n0}C(0)\rho_{10}+\delta_{m1}\delta_{n1}|C(0)|^2\rho_{11}. \label{eq:(63)}
\end{align}
If we introduce a phase shifter to eliminate the phase of $C(0)$, the converted signal state is as follows:
\begin{align}
\rho_{s}(L)=
\begin{bmatrix}
\rho_{00}+(1-\eta_{d})\rho_{11} & \sqrt{\eta_{d}}\rho_{01}\\
\sqrt{\eta_{d}}\rho_{10} & \eta_d\rho_{11}
\end{bmatrix}. \label{eq:(64)}
\end{align}
The qubit encoded in the photon-number DOF of the probe field is perfectly preserved in the converted signal field when the CE reaches $100\%$. For the frequency up-conversion, the result can be obtained by interchanging the $p$ and $s$ symbols and replacing $C(0)$ with $B(0)$ and $\eta_d$ with $\eta_u$. The preservation of entanglement between the single-rail qubits after applying the QFC scheme is demonstrated in Sec. \ref{sec:entanglement retention} for the most general N-qubit case. The results indicate that the QI can be fully preserved using the N-qubit QFC scheme if we eliminate the phases of the output fields and achieve unity CE for each conversion channel.


\subsection{Path-Encoded Qubit} \label{subsec:path encoded qubit}

The path DOF can be harnessed to prepare high-dimensional photonic quantum states \cite{Path Encoded Qubits source, Path Encoded Qubits QKD}, and it exhibits excellent compatibility with photonic integrated quantum circuits \cite{Photonic Integrated Circuits Nondeterministic, Photonic Integrated Circuits Deterministic}. We first consider a diamond-type QFC scheme that involves two separate atomic ensembles: the up-ensemble and the down-ensemble, as depicted in Fig. \ref{fig:path encoded qubit}(a). The logical basis for path encoding is defined as $\ket{0}=\ket{1}_D\ket{0}_U$ and $\ket{1}=\ket{0}_D\ket{1}_U$, where $\ket{i}_U$ and $\ket{i}_D$ indicate the number states of the up and down spatial modes, respectively. Thus, the qubit is also referred to as the dual-rail qubit \cite{Single-Rail Qubit Passive Quantum Computing, Single-Rail and Dual-Rail Interface}. The two QFC processes must be considered as a whole since entanglement can be shared between the two spatial modes of the electromagnetic field. We employ a similar approach to that in Sec. \ref{sec:converted signal state}, but with higher dimensions comprising up- and down-paths, where $\rho_f=U\rho_i U^{\dagger}$ and $\rho_i=\rho_s(0)\otimes\rho_p(0)\otimes\rho_R$. Here, $U$ represents the evolution operator of the entire system, and $\rho_{s(p)}(0)$ comprises up- and down-paths.

Consider a frequency down-conversion from the probe to the signal, as depicted in Fig. \ref{fig:path encoded qubit}(a), the input signal field is assumed to be in the vacuum state. Note that in this and the next sections, the subscripts $D$ and $U$ stand for down- and up-ensembles, respectively. Expanding the converted signal density operator with respect to the combined number basis $\ket{i_D j_U}\equiv \ket{i}_D\ket{j}_U$, the density matrix element is as follows:
\begin{align}
&\rho_{m_D m_U n_D n_U}^s(L)=\bra{m_D m_U}\rho_s(L)\ket{n_D n_U}\notag\\
&=tr\left\{U^{\dagger}(\ket{n_D n_U}\bra{m_D m_U}\otimes I_p\otimes I_R)U\rho_i\right\}. \label{eq:(65)}
\end{align}
We can once again represent $\ket{n_D n_U}\bra{m_D m_U}$ as the sum of multiplication of the initial ladder operators. Following a similar approach to that in Sec. \ref{sec:converted signal state}, the density matrix element of the converted signal field can be obtained as follows:
\begin{align}
\rho_{m_D m_U n_D n_U}^s(L)=\sum_{l_D, l_U=0}^{\infty}\chi&_{m_D m_U n_D n_U l_D l_U}\notag\\
tr_p\big\{[C_D^*(0)\widetilde{a}_{p,D}^{\dagger}(0)]^{l_D+n_D}&[C_U^*(0)\widetilde{a}_{p,U}^{\dagger}(0)]^{l_U+n_U}\notag\\
[C_D(0)\widetilde{a}_{p,D}(0)]^{l_D+m_D}&[C_U(0)\widetilde{a}_{p,U}(0)]^{l_U+m_U}\rho_p(0)\big\}\notag\\
\equiv\sum_{l_D, l_U=0}^{\infty}\chi_{m_D m_U n_D n_U l_D l_U}&tr_p\{\hat{O}\rho_p(0)\}, \label{eq:(66)}
\end{align}
where $\chi_{m_D m_U n_D n_U l_D l_U}\equiv\frac{(-1)^{l_D}(-1)^{l_U}}{l_D!l_U!}\frac{1}{\sqrt{m_D!m_U!}}\frac{1}{\sqrt{n_D!n_U!}}$. By utilizing the above expression, we can obtain the converted signal state for any input probe state while considering the entire system. Now, let us consider the case of the input probe field with an arbitrary one-qubit state using path encoding, and the density operator of the input probe state is given by
\begin{align}
\rho_p(0)=&\rho_{00}\ket{1_D 0_U}\bra{1_D 0_U}+\rho_{01}\ket{1_D 0_U}\bra{0_D 1_U}\notag\\
+&\rho_{10}\ket{0_D 1_U}\bra{1_D 0_U}+\rho_{11}\ket{0_D 1_U}\bra{0_D 1_U}. \label{eq:(67)}
\end{align}
Here, as defined earlier, $\ket{1_D 0_U}=\ket{0}$ and $\ket{0_D 1_U}=\ket{1}$. The converted signal state can be obtained using Eq. (\ref{eq:(66)}), and the density matrix element is as follows:
\begin{align}
\rho_{m_D m_U n_D n_U}^s(L)=\sum_{l_D, l_U=0}^{\infty}&\chi_{m_D m_U n_D n_U l_D l_U}\notag\\
\left[\rho_{00}\bra{1_D 0_U}\hat{O}\ket{1_D 0_U}+\right.&\rho_{01}\bra{0_D 1_U}\hat{O}\ket{1_D 0_U}\notag\\
+\rho_{10}\bra{1_D 0_U}\hat{O}\ket{0_D 1_U}+&\left.\rho_{11}\bra{0_D 1_U}\hat{O}\ket{0_D 1_U}\right], \label{eq:(68)}
\end{align}
where each expectation value term can be calculated by separating the down and up Hilbert space components. The converted density operator is then obtained as follows:
\begin{align}
&\rho_s(L)\notag\\
&=\rho_{00}|C_D|^2\ket{1_D 0_U}\bra{1_D 0_U}+\rho_{01}C_D C_U^*\ket{1_D 0_U}\bra{0_D 1_U}\notag\\
&+\rho_{10}C_D^* C_U\ket{0_D 1_U}\bra{1_D 0_U}+\rho_{11}|C_U|^2\ket{0_D 1_U}\bra{0_D 1_U}\notag\\
&+\left[\rho_{00}(1-|C_D|^2)+\rho_{11}(1-|C_U|^2)\right]\ket{0_D 0_U}\bra{0_D 0_U}. \label{eq:(69)}
\end{align}
If we introduce phase shifters at the output of the up- and down-ensembles to eliminate the phase of $C_U(0)$ and $C_D(0)$, and omit the vacuum term, the converted density operator using the logical basis representation is as follows:
\begin{align}
\rho_s(L)=
\begin{bmatrix}
\eta_D\rho_{00} & \sqrt{\eta_D\eta_U}\rho_{01}\\
\sqrt{\eta_U\eta_D}\rho_{10} & \eta_U\rho_{11}
\end{bmatrix}, \label{eq:(70)}
\end{align}
where the qubit encoded in the path DOF is perfectly preserved when both conversion processes achieve $100\%$ CE.
\FigFive

Suppose we input a path-encoded qubit in a superposition state $\frac{1}{\sqrt{2}}(\ket{1_D 0_U}+\ket{0_D 1_U})$. The dual-rail encoding state can also be identified as a maximally entangled state (Bell state) $\ket{\Psi^+}$ if we consider the up- and down-paths as the QI carriers of two single-rail qubits (i.e., qubit $D$ and $U$). We can then calculate the converted two-qubit state from Eq. (\ref{eq:(70)}), which yields the following result:
\begin{align}
&\rho_s(L)\notag\\
&=\frac{\eta_D}{2}\ket{1}_D\ket{0}_U\bra{1}_D\bra{0}_U+\frac{\sqrt{\eta_D\eta_U}}{2}\ket{1}_D\ket{0}_U\bra{0}_D\bra{1}_U\notag\\
&+\frac{\sqrt{\eta_U\eta_D}}{2}\ket{0}_D\ket{1}_U\bra{1}_D\bra{0}_U+\frac{\eta_U}{2}\ket{0}_D\ket{1}_U\bra{0}_D\bra{1}_U, \label{eq:(71)}
\end{align}
where if both CEs reach $100\%$, the converted state is exactly $\ket{\Psi^+}$. Entanglement between the two single-rail qubits $D$ and $U$ has been perfectly retained. The telecom-band QFC on such a single-rail entangled state can be applied in the DLCZ protocol \cite{DLCZ Protocol, DLCZ Protocol Realization} as depicted in Fig. \ref{fig:path encoded qubit}(b), enhancing the efficiency of quantum communication. For the frequency up-conversion, the result can be obtained by interchanging the $p$ and $s$ symbols and replacing $C_{U(D)}(0)$ with $B_{U(D)}(0)$. The preservation of entanglement between the path-encoded qubits after applying the QFC scheme is demonstrated in Sec. \ref{sec:entanglement retention} for the most general N-qubit case. The QI can also be completely preserved using the N-qubit QFC scheme if we eliminate the phases of the output fields and reach unity CE for each conversion channel.


\subsection{Polarization-Encoded Qubit} \label{subsec:polarization encoded qubit}

The polarization DOF has been widely adopted for qubit encoding. This is attributed to several factors, including the ease of obtaining polarization-entangled photon pair sources \cite{Polarization-Entangled Biphoton, Single-Photon Sources-1, Single-Photon Sources-2} and the straightforward manipulation and projection measurements of qubits using basic optical elements \cite{Linear Optical Quantum Computing-1, Polarization Encoded Qubits CNOT Gate-1, Polarization Encoded Qubits CNOT Gate-2}. Here, we theoretically demonstrate that the diamond-type QFC successfully performs frequency conversion for polarization-encoded qubits. We define the logical basis as $\ket{0}=\ket{1_H 0_V}$ and $\ket{1}=\ket{0_H 1_V}$, where $H$ and $V$ represent horizontal and vertical polarization, respectively. Considering the configuration presented in Fig. \ref{fig:polarization encoded qubit}, we input a polarization-encoded single photon with an arbitrary one-qubit state as follows:
\begin{align}
\rho_{in}=&\rho_{00}\ket{0}\bra{0}+\rho_{01}\ket{0}\bra{1}\notag\\
+&\rho_{10}\ket{1}\bra{0}+\rho_{11}\ket{1}\bra{1}. \label{eq:(72)}
\end{align}
The polarization beam splitter (PBS) separates different polarization components along distinct paths \cite{Beam Splitter}, thereby converting qubit information from polarization to the path DOF, resulting in a path-encoded qubit along the up- and down-paths. To achieve polarization-stable QFC and ensure simultaneous optimization of QFCs on both polarization components, we propose the use of two QFC systems instead of one. In this scenario, we specifically focus on frequency down-conversion from the probe to the signal. The probe density operator after passing through the PBS is as follows:
\begin{align}
\rho_p(0)=&\rho_{00}\ket{1_D 0_U}\bra{1_D 0_U}+i\rho_{01}\ket{1_D 0_U}\bra{0_D 1_U}\notag\\
-&i\rho_{10}\ket{0_D 1_U}\bra{1_D 0_U}+\rho_{11}\ket{0_D 1_U}\bra{0_D 1_U}, \label{eq:(73)}
\end{align}
where the field along the lower path is horizontally polarized, while the field along the upper path is vertically polarized. Quarter-wave plates (QWPs) are used on each side to convert the fields passing through the up- and down-paths into circular polarization before entering the respective diamond-type atomic ensembles. In this context, we specifically chose circular polarization to align with the requirements of the optimized QFC scheme discussed in Section \ref{subsec:conversion efficiency optimization}. Note that both atomic systems must be appropriately configured for their respective QFC schemes.

\FigSix

After the first set of QWPs, the QFC process mirrors that of the path-encoded qubit discussed in Section \ref{subsec:path encoded qubit}. Subsequently, the second set of QWPs transforms the circularly polarized up- and down-output fields back to their original vertical and horizontal polarizations, respectively. Once the phase shifters neutralize the phases of $C_U(0)$ and $C_D(0)$, as well as the phase changes from the previous and subsequent PBSs, the resulting converted signal state is as follows:
\begin{align}
&\rho_s(L)\notag\\
&=\eta_D\rho_{00}\ket{1_D 0_U}\bra{1_D 0_U}-i\sqrt{\eta_D\eta_U}\rho_{01}\ket{1_D 0_U}\bra{0_D 1_U}\notag\\
&+i\sqrt{\eta_U\eta_D}\rho_{10}\ket{0_D 1_U}\bra{1_D 0_U}+\eta_U\rho_{11}\ket{0_D 1_U}\bra{0_D 1_U}, \label{eq:(74)}
\end{align}
where the vacuum term $\ket{0_D 0_U}\bra{0_D 0_U}$ is omitted. The second PBS combines the up- and down-fields with distinct spatial modes into a single path (another output port is in the vacuum state). The density operator of the output signal field after the PBS using the logical basis representation is as follows:
\begin{align}
\rho_{out}=
\begin{bmatrix}
\eta_D\rho_{00} & \sqrt{\eta_D\eta_U}\rho_{01}\\
\sqrt{\eta_U\eta_D}\rho_{10} & \eta_U\rho_{11}
\end{bmatrix}, \label{eq:(75)}
\end{align}
where the qubit encoded in the polarization DOF of the input field $\rho_{in}$ has been perfectly preserved, reaching unity fidelity when both QFC processes achieve $100\%$ CE. The result for frequency up-conversion can be obtained by interchanging the $p$ and $s$ symbols and replacing $C_{U(D)}(0)$ with $B_{U(D)}(0)$. The preservation of entanglement between the polarization-encoded qubits after applying the QFC scheme is demonstrated in Section \ref{sec:entanglement retention} for the most general N-qubit case, and it is consistent with the preservation observed in the path-encoded case. Perfect preservation is achievable by eliminating the phases of the output fields and achieving unity CE for each conversion channel.


\section{Entanglement Retention} \label{sec:entanglement retention}

\subsection{Multiple Qubits} \label{subsec:N-Qubit QFC}

In this section, we extend the system to implement QFC involving an arbitrary number of qubits, denoted as an N-qubit system. These qubits may be encoded in photon-number, path, or polarization DOFs, and quantum entanglement can exist among them. We first explore a diamond-type QFC system that includes $N$ separated atomic ensembles, denoted as $A_i$ with $i\in\{1,2,\ldots ,N\}$. Each atomic ensemble $A_i$ has probe and signal input fields, denoted as $p_i(0)$ and $s_i(0)$, respectively. For the down-conversion case, all signal fields initially exist in the vacuum states, and the combined state is denoted as $\rho_s(0)=\ket{0_10_2\ldots 0_N}\bra{0_10_2\ldots 0_N}$. Since the input probe fields of the combined system must be treated as a whole, we consider the combined density operator $\rho_p(0)$ of all probe fields as the input state. Using the reduced-density-operator approach, we derive the combined density operator of all output signal fields $\rho_s(L)$. Here, we expand the output signal state as follows:
\begin{align}
\rho_s&(L)=\sum_{m_1=0}^{\infty}\ldots\sum_{m_N=0}^{\infty}\sum_{n_1=0}^{\infty}\ldots\sum_{n_N=0}^{\infty}\notag\\
&\rho_{m_1\ldots m_N, n_1\ldots n_N}^s(L)\ket{m_1\ldots m_N}\bra{n_1\ldots n_N}. \label{eq:(76)}
\end{align}
Following a similar derivation as that in Sec. \ref{sec:converted signal state}, the density matrix element of the converted signal field can be expressed as follows:
\begin{align}
\rho&_{m_1\ldots m_N, n_1\ldots n_N}^s(L)\notag\\
=&tr\left\{U^{\dagger}\left(\ket{n_1\ldots n_N}\bra{m_1\ldots m_N}\otimes I_p\otimes I_R\right)U\rho_i\right\}\notag\\
=&\sum_{l_1=0}^{\infty}\ldots\sum_{l_N=0}^{\infty}\chi_{m_1\ldots m_Nn_1\ldots n_Nl_1\ldots l_N}\notag\\
&tr\Bigg\{\prod_{j=1}^{N}\big[\widetilde{a}_{s, j}^{\dagger}(L)\big]^{n_j+l_j}\prod_{k=1}^{N}\big[\widetilde{a}_{s, k}(L)\big]^{m_k+l_k}\rho_i\Bigg\}, \label{eq:(77)}
\end{align}
where the term $\chi_{m_1\ldots m_Nn_1\ldots n_Nl_1\ldots l_N}$ is defined as $\frac{(-1)^{l_1+\ldots+l_N}}{l_1!\ldots l_N!} \frac{1}{\sqrt{m_1!\ldots m_N!n_1!\ldots n_N!}}$. As the output ladder operator of the QFC for each atomic ensemble follows the same form as in the single-mode version of Eq. (\ref{eq:(17)}), the trace term in Eq. (\ref{eq:(77)}) can be simplified using a similar approach as outlined in Appendix \ref{sec:Appendix F}. Consequently, the density matrix element for the converted signal field can be derived, and its explicit form is provided below:
\begin{align}
&\rho_{m_1\ldots m_N, n_1\ldots n_N}^s(L)\notag\\
&=\sum_{l_1=0}^{\infty}\ldots\sum_{l_N=0}^{\infty}\chi_{m_1\ldots m_Nn_1\ldots n_Nl_1\ldots l_N}tr_p\Bigg\{\prod_{j=1}^{N}\notag\\
&\big[C_j^*\widetilde{a}_{p, j}^{\dagger}(0)\big]^{n_j+l_j}\prod_{k=1}^{N}\big[C_k\widetilde{a}_{p, k}(0)\big]^{m_k+l_k}\rho_p(0)\Bigg\}. \label{eq:(78)}
\end{align}

Consider an N-qubit input state $\rho_p(0)$ which is physically implemented by N spatial modes of the electromagnetic field, with each qubit encoded in the photon-number DOF of its corresponding spatial mode. The $i$th qubit (or $i$th field mode) is sent into the $A_i$ atomic ensemble, undergoing frequency conversion through the diamond-type QFC process. The density operator of the input probe field can be expressed as follows:
\begin{align}
\rho_p&(0)=\sum_{m_1=0}^{1}\ldots\sum_{m_N=0}^{1}\sum_{n_1=0}^{1}\ldots\sum_{n_N=0}^{1}\notag\\
&\rho_{m_1\ldots m_N, n_1\ldots n_N}^p(0)\ket{m_1\ldots m_N}\bra{n_1\ldots n_N}. \label{eq:(79)}
\end{align}
By utilizing the separability between operators associated with distinct atomic ensembles, we have derived the density matrix element for the converted signal field. The overall expression for the density matrix element of the converted signal field is as follows:
\begin{align}
\rho&_{m_1\ldots m_N, n_1\ldots n_N}^s(L)\notag\\
=&\sum_{q_1=0}^{1}\ldots\sum_{q_N=0}^{1}\sum_{r_1=0}^{1}\ldots\sum_{r_N=0}^{1}\rho_{q_1\ldots q_N, r_1\ldots r_N}^p(0)\notag\\
&\prod_{j=1}^N\big\{\delta_{m_jn_j,00}\left[\delta_{q_jr_j,00}+\delta_{q_jr_j,11}(1-\eta_j)\right]\notag\\
&+\delta_{m_jn_j,01}\delta_{q_jr_j,01}C_j^*+\delta_{m_jn_j,10}\delta_{q_jr_j,10}C_j\notag\\
&+\delta_{m_jn_j,11}\delta_{q_jr_j,11}\eta_j\big\}, \label{eq:(80)}
\end{align}
where we define $\delta_{ab,cd}=\delta_{ac}\delta_{bd}$, and $\eta_j$ represents the CE of the diamond-type QFC through atomic ensemble $A_j$. If the CE for each QFC reaches $100\%$, and the phase of each $C_j$ has been eliminated by the phase shifter, we can deduce that
\begin{align}
\rho&_{m_1\ldots m_N, n_1\ldots n_N}^s(L)=\rho_{m_1\ldots m_N, n_1\ldots n_N}^p(0), \label{eq:(81)}
\end{align}
which leads to $\rho_s(L)=\rho_p(0)$ and thereby confirms that the diamond-type QFC scheme has perfectly converted the single-rail encoded N-qubit state from the input probe field to the output signal field.

The discussion for the path-encoded N-qubit QFC scheme parallels the single-rail case. Initially, all atomic ensembles are grouped into pairs, where the $i$th pair consists of the $A_i$ and $B_i$ ensembles. Each pair is dedicated to converting a path-encoded qubit, denoted as $Q_i$. The basis for the path-encoded single qubit is given by $\ket{0_{Q_j}}=\ket{0_{A_j}1_{B_j}}$ and $\ket{1_{Q_j}}=\ket{1_{A_j}0_{B_j}}$, while the N-qubit basis can be represented as $\ket{s_{Q_1}\ldots s_{Q_N}}=\ket{(s_{Q_1})_{A_1}(1-s_{Q_1})_{B_1}\ldots(s_{Q_N})_{A_N}(1-s_{Q_N})_{B_N}}$, where $s_{Q_1}, \ldots, s_{Q_N}\in\{0,1\}$. Expressing the density operator of the N-qubit input probe field is as follows:
\begin{align}
\rho_p&(0)=\sum_{s_{Q_1}=0}^{1}\ldots\sum_{s_{Q_N}=0}^{1}\sum_{t_{Q_1}=0}^{1}\ldots\sum_{t_{Q_N}=0}^{1}\notag\\
&\rho_{s_{Q_1}\ldots s_{Q_N}, t_{Q_1}\ldots t_{Q_N}}^p(0)\ket{s_{Q_1}\ldots s_{Q_N}}\bra{t_{Q_1}\ldots t_{Q_N}}. \label{eq:(82)}
\end{align}
If we express the dual-rail basis $\ket{s_{Q_j}}$ (and $\bra{t_{Q_j}}$) using the single-rail logical basis, $\ket{s_{Q_j}}=\ket{(s_{Q_j})_{A_j}(1-s_{Q_j})_{B_j}}$, the input state in Eq. (\ref{eq:(82)}) can be regarded as a special case of Eq. (\ref{eq:(79)}). The conversion process can then be described based on the previous single-rail discussion. The converted signal density matrix element would take the form as in Eq. (\ref{eq:(80)}). The path-encoded N-qubit state is perfectly preserved if the CEs for all conversion processes reach $100\%$, and if all the phases have been eliminated. For the QFC scheme of the polarization-encoded N-qubit state, we can use the method introduced in Fig. \ref{fig:polarization encoded qubit} to map the polarization-encoded state into the path-encoded state, and then map back into the polarization-encoded state after all the conversion processes are finished. Hence, the polarization-encoded N-qubit state can also reach unity fidelity if each QFC has perfect CE and the phase has been eliminated. These results suggest that the N-qubit QFC scheme with diamond-type four-wave mixing can actually function as an N-qubit quantum interface for single-rail-, path-, and polarization-encoded qubits.


\subsection{Polarization-Entangled EPR Pairs} \label{subsec:EPR Pairs}

Here, we provide a detailed analysis on the retention of EPR pairs after the diamond-type QFC. Consider a pair of polarization entangled photons in the Bell state $\ket{\Phi^+}=\frac{1}{\sqrt{2}}(\ket{0_10_2}+\ket{1_11_2})$ with a near-infrared wavelength; the logical basis is defined the same as in Sec. \ref{subsec:polarization encoded qubit}. As depicted in Fig. \ref{fig:EPR pairs}(a), the horizontal and vertical components of polarization qubit $Q_i$ are separated and sent into atomic ensembles $B_i$ and $A_i$, respectively. The combined state of the input probe fields can be expressed as follows:
\begin{align}
\ket{\psi_p(0)}=\frac{1}{\sqrt{2}}(\ket{0_{A_1}1_{B_1}0_{A_2}1_{B_2}}+\ket{1_{A_1}0_{B_1}1_{A_2}0_{B_2}}), \label{eq:(83)}
\end{align}
where the phase shifters are employed to eliminate the phases of the output signal fields. On each side, the different polarization components of the converted photon are combined using a PBS. The converted qubits are well-suited for long-distance transmission through optical fibers. In practical quantum communication, where coincidence detections are required, the original converted state $\rho_s(L)$ can be post-projected onto the basis with one photon on each side (i.e., $\ket{0_{A_1}1_{B_1}0_{A_2}1_{B_2}}$, $\ket{0_{A_1}1_{B_1}1_{A_2}0_{B_2}}$, $\ket{1_{A_1}0_{B_1}0_{A_2}1_{B_2}}$, and $\ket{1_{A_1}0_{B_1}1_{A_2}0_{B_2}}$). By using Eq. (\ref{eq:(80)}), we can calculate the converted density matrix elements corresponding to the post-projection basis. The non-zero terms are as follows:
\begin{align}
\rho^s_{0101, 0101}(L)&=\frac{1}{2}\eta_{B_1}\eta_{B_2}, \label{eq:(84)}\\
\rho^s_{1010, 1010}(L)&=\frac{1}{2}\eta_{A_1}\eta_{A_2}, \label{eq:(85)}\\
\rho^s_{0101, 1010}(L)&=\rho^s_{1010, 0101}(L)=\frac{1}{2}\sqrt{\eta_{A_1}\eta_{A_2}\eta_{B_1}\eta_{B_2}}. \label{eq:(86)}
\end{align}
The converted density operator under post-selection can then be expressed as follows:
\begin{align}
\rho&_{s, post}(L)\notag\\
=&\frac{\bar{\eta}_B^2}{\bar{\eta}_A^2+\bar{\eta}_B^2}\ket{0_10_2}\bra{0_10_2}+\frac{\bar{\eta}_A\bar{\eta}_B}{\bar{\eta}_A^2+\bar{\eta}_B^2}\ket{0_10_2}\bra{1_11_2}\notag\\
+&\frac{\bar{\eta}_A\bar{\eta}_B}{\bar{\eta}_A^2+\bar{\eta}_B^2}\ket{1_11_2}\bra{0_10_2}+\frac{\bar{\eta}_A^2}{\bar{\eta}_A^2+\bar{\eta}_B^2}\ket{1_11_2}\bra{1_11_2}, \label{eq:(87)}
\end{align}
where we define $\bar{\eta}_A=\sqrt{\eta_{A_1}\eta_{A_2}}$ and $\bar{\eta}_B=\sqrt{\eta_{B_1}\eta_{B_2}}$. The coincidence detection probability can be obtained as follows:
\begin{align}
P_c=\sum_i\bra{i}\rho_s(L)\ket{i}=\frac{1}{2}(\bar{\eta}_A^2+\bar{\eta}_B^2), \label{eq:(88)}
\end{align}
where $\ket{i}$ represents the post-projection basis. $P_c$ is also the success rate of photon pair transmission, which is directly associated with the efficiency of QI transmission. The fidelity between the post-selected output signal field and the input probe field can be obtained as follows:
\begin{align}
F=\frac{\bar{\eta}_A+\bar{\eta}_B}{\sqrt{2(\bar{\eta}_A^2+\bar{\eta}_B^2)}}. \label{eq:(89)}
\end{align}
The post-selected fidelity is illustrated in Fig. \ref{fig:EPR pairs}(b). The fidelity increases as $\bar{\eta}_A$ approaches $\bar{\eta}_B$, and reaches unity when $\bar{\eta}_A=\bar{\eta}_B$. While the post-selected converted state closely resembles $\ket{\Phi^+}$ for low CEs, the transmission rate ($P_c$) of photon pairs diminishes in such cases, resulting in reduced QI transmission efficiency.

\FigSeven

In order to assess the retention of quantum entanglement, we employ Bell's inequality \cite{Bell Nonlocality, Bell Inequality}. Specifically, we utilize the Clauser-Horne-Shimony-Holt (CHSH) inequality \cite{Bell Nonlocality, CHSH Inequality}, which is a specific type of Bell's inequality. Consider a Bell test between Alice and Bob, as depicted in Fig. \ref{fig:EPR pairs}(a). The converted polarization qubits, denoted as $Q_1'$ and $Q_2'$, are sent to Alice and Bob, respectively. The CHSH inequality is expressed as
\begin{align}
|S|\leq2, \label{eq:(90)}
\end{align}
where the Bell operator $S$ is defined as
\begin{align}
S=\braket{\hat{A}_0\hat{B}_0}+\braket{\hat{A}_0\hat{B}_1}+\braket{\hat{A}_1\hat{B}_0}-\braket{\hat{A}_1\hat{B}_1}. \label{eq:(91)}
\end{align}
In order to give the maximum violation of CHSH inequality, we consider the following local observables:
\begin{align}
&\hat{A}_0=\hat{\sigma}_z, \label{eq:(92)}\\
&\hat{A}_1=\hat{\sigma}_x, \label{eq:(93)}\\
&\hat{B}_0=\frac{1}{\sqrt{2}}(\hat{\sigma}_z+\hat{\sigma}_x), \label{eq:(94)}\\
&\hat{B}_1=\frac{1}{\sqrt{2}}(\hat{\sigma}_z-\hat{\sigma}_x). \label{eq:(95)}
\end{align}
Here, Alice opts to perform measurements with test angles of $0^{\circ}$ or $45^{\circ}$, while Bob selects test angles of $22.5^{\circ}$ or $67.5^{\circ}$. The measurement at $0^{\circ}$ corresponds to the projection onto the orthogonal qubit basis $\ket{0_1}$ and $\ket{1_1}$. Coincidences between Alice and Bob are recorded in the Bell test. The Bell operator for the converted state and the local observables (\ref{eq:(92)})--(\ref{eq:(95)}) can be determined as follows:
\begin{align}
S=\frac{\left(\bar{\eta}_A+\bar{\eta}_B\right)^2}{\bar{\eta}_A^2+\bar{\eta}_B^2}\sqrt{2}=2\sqrt{2}F^2. \label{eq:(96)}
\end{align}
To violate the CHSH inequality, the post-selected fidelity needs to be greater than $2^{-1/4}$, which is $F>2^{-1/4}\approx84.1\%$. As depicted in Fig. \ref{fig:EPR pairs}(b), conversion processes with a combination of CEs in the range of $F>2^{-1/4}$ exhibit quantum nonlocality between the converted photons, defying interpretation through local hidden-variable theories \cite{EPR Paradox, Loophole-Free Bell Inequality Violation, Bell’s Theorem and Locally Mediated Models}. The nonlocality between the original qubits $Q_1$ and $Q_2$ has been partially retained after the QFC. For conversion processes with the same fidelity, their post-selected density operators present the following two possibilities:
\begin{align}
\rho_{s, post}(L)=&F^2\ket{\Phi^+}\bra{\Phi^+}+(1-F^2)\ket{\Phi^-}\bra{\Phi^-}\notag\\
\pm&F\sqrt{1-F^2}\left(\ket{\Phi^+}\bra{\Phi^-}+\ket{\Phi^-}\bra{\Phi^+}\right). \label{eq:(97)}
\end{align}
In Fig. \ref{fig:EPR pairs}(b), the fidelity surface and the fidelity plane intersect at two lines. The conversion processes associated with these two lines correspond to the two different density operators as in Eq. (\ref{eq:(97)}). The second density operator in Eq. (\ref{eq:(97)}) with $F=2^{-1/4}$ is illustrated in Fig. \ref{fig:EPR pairs}(c), where the probability of the post-selected output state remaining in the Bell state $\ket{\Phi^+}$ is $2^{-1/2}\approx70.7\%$. Similar results can be obtained for other Bell states. These findings suggest that the QFC scheme can effectively retain the entanglement between the polarization entangled photons with high fidelity if the four QFC systems are appropriately adjusted, ensuring that all CEs are sufficiently close.


\section{Conclusion} \label{sec:conclusion}

This study provides a theoretical analysis of the QFC scheme employing a diamond-type energy level configuration within a rubidium atomic ensemble. Using the Heisenberg-Langevin approach, we derive the general forms of the field operators for the probe and signal fields. The model, addressing the absorption of the coupling field, precisely solves the coupling field MSE and introduces Magnus expansion into the coupled equations. Importantly, physical parameters such as CE and transmittance remain unaffected by vacuum field noise, enabling the system to achieve high-purity QFC.

We optimize transition schemes for real-world applications systematically. Through an extensive parameter scan, we identify optimal parameters that maximize CE at different ODs. The CE increases with higher ODs and can physically approach $100\%$, surpassing $90\%$ at an OD of approximately 240 for the telecom E-band, and reaching $80\%$ at an OD of 700 for the C-band. Emphasizing the importance of considering coupling field absorption in the high CE regime, we demonstrate its significance through comparison with a nonabsorbing model, highlighting the crucial impact on practical applications.

Various quantum properties within the diamond-type QFC system are explored. The derivations of the quadrature variances for the converted field reveal that, upon eliminating the phase of the output field, the quadrature variances mirror those of the input field when the CE reaches $100\%$. Additionally, a detailed analysis of the converted quadrature variances for specific input states, including the n-photon Fock state, coherent state, and squeezed coherent state, is conducted.

We derive the exact form of the density operator for the converted field using the reduced-density-operator approach. For the case of an input field in the single-photon Fock state $\ket{1}$, the probability of the output field retaining the Fock state $\ket{1}$ corresponds to the CE. Similarly, for a coherent input state $\ket{\beta}$, the resulting converted state remains a coherent state $\ket{C(0)\beta}$. In the case where we eliminate the phase of the output field, the fidelity between input and converted states approaching perfection as CE approaches $100\%$.

The diamond-type QFC scheme exhibits exceptional capability in preserving QI encoded in photon-number, path, and polarization DOFs. Highly preservation of quantum states for single-rail, path, and polarization encoded qubits is demonstrated for sufficiently high CE, achieving unity fidelity at $100\%$ CE. The extension of the QFC system enables implementation on an arbitrary number of entangled qubits, maintaining the density operator of the converted N-qubit entangled state for perfect CE. Theoretical indications suggest the diamond-type QFC scheme can indeed serve as a robust quantum interface, facilitating frequency conversion while preserving quantum states.

In conclusion, a customizable diamond-type QFC scheme in a rubidium atomic ensemble proves highly efficient, bridging the gap between near-infrared and telecom E-band or C-band. The QFC scheme highly preserved the QI carried by the photons. The CE and fidelity of the quantum state are unaffected by vacuum field noise, positions the QFC system to deliver high-purity converted photons. This quantum interface holds immense potential for connecting quantum memory and processing systems emitting in the near-infrared range to the telecom wavelength, laying the foundation for a robust quantum communication network in the future.


\section*{ACKNOWLEDGMENTS}

This work was supported by the National Science and Technology Council of Taiwan under Grants No. 111-2112-M-006-027, No. 112-2112-M-006-034, and No. 112-2119-M-007-007. We also acknowledge support from the Center for Quantum Science and Technology (CQST) within the framework of the Higher Education Sprout Project by the Ministry of Education (MOE) in Taiwan.


\appendix

\section{HEISENBERG-LANGEVIN EQUATIONS AND RELAXATION TERMS} \label{sec:Appendix A}

A dissipative system, affected by the influence of a background reservoir over time, can be described by the HLE. The derivation of the HLE initiates from the Heisenberg equation, which can be expressed under the SVA basis as follows:
\begin{align}
&\frac{\partial\hat{\sigma}_{ij}}{\partial t}=\frac{i}{\hbar}\big[\hat{H}_{tot},\hat{\sigma}_{ij}\big], \label{eq:(A1)}\\
&\hat{H}_{tot}=\hat{H}_{S}+\hat{H}_{R}+\hat{H}_{SR}, \label{eq:(A2)}\\
&\hat{H}_{S}=\int_0^L \frac{N}{L}\sum_{ij}\hat{H}_{ij}(z,t)\hat{\sigma}_{ij}(z,t) dz, \label{eq:(A3)}
\end{align}
where $\hat{H}_{ij}(z,t)$ represents the matrix element of the averaged single-atom Hamiltonian at position $z$ under the atomic basis representation; $\hat{H}_{S}$, $\hat{H}_{R}$, and $\hat{H}_{SR}$ are the Hamiltonians for the system, reservoir, and system-reservoir interaction, respectively. Further derivation enables the terms associated with the reservoir in Eq. (\ref{eq:(A1)}) to be decomposed into relaxation and fluctuation components as follows:
\begin{align}
\frac{\partial\hat{\sigma}_{ij}}{\partial t}=\frac{i}{\hbar}\big[\hat{H}_S,\hat{\sigma}_{ij}\big]+\hat{R}_{ij}+\hat{F}_{ij}, \label{eq:(A4)}
\end{align}
where $\hat{R}_{ij}$ describes the dissipation process and $\hat{F}_{ij}$ is the Langevin noise operator from the background reservoir. This equation is known as the HLE. By substituting in the Hamiltonian of the whole system $\hat{H}_S$, the HLE can be rearranged into the following form:
\begin{align}
\frac{\partial}{\partial t}\hat{\sigma}_{ij}&=\frac{i}{\hbar}\sum_{m}(\hat{H}_{im}^*\hat{\sigma}_{mj}-\hat{\sigma}_{im}\hat{H}_{mj}^*)+\hat{R}_{ij}+\hat{F}_{ij}. \label{eq:(A5)}
\end{align}
This expression represents a slice of the HLE at the position $z$. By utilizing the Hermitian property of the single-atom Hamiltonian, we can collect all of the HLEs into a matrix form as follows:
\begin{align}
\frac{\partial}{\partial t}\hat{\sigma}(z,t)=\frac{i}{\hbar}\big[\hat{H}^T(z,t),\hat{\sigma}(z,t)\big]+\hat{R}(z,t)+\hat{F}(z,t). \label{eq:(A6)}
\end{align}
In this matrix equation, $\hat{H}^T(z,t)$ must be transformed into a matrix form based on the averaged single-atom basis at position $z$. The equation derived from the $ij$ matrix element of Eq. (\ref{eq:(A6)}) corresponds to the HLE associated with $\hat{\sigma}_{ij}$. The system-reservoir interaction of the diamond-type QFC system can be approximated as a Markovian process; thus, the relaxation terms that adhere to the selection rules of orbital angular momentum $\Delta l=\pm 1$ are as follows:
\begin{align}
&\begin{bmatrix}
\begin{smallmatrix}
\Gamma_{21}\hat{\sigma}_{22}+\Gamma_{31}\hat{\sigma}_{33} & -\frac{1}{2}\gamma_{21}\hat{\sigma}_{12} & -\frac{1}{2}\gamma_{31}\hat{\sigma}_{13} & -\frac{1}{2}\gamma_{41}\hat{\sigma}_{14}\\
-\frac{1}{2}\gamma_{21}\hat{\sigma}_{21} & \Gamma_{42}\hat{\sigma}_{44}-\Gamma_{21}\hat{\sigma}_{22} & -\frac{1}{2}\gamma_{32}\hat{\sigma}_{23} & -\frac{1}{2}\gamma_{42}\hat{\sigma}_{24}\\
-\frac{1}{2}\gamma_{31}\hat{\sigma}_{31} & -\frac{1}{2}\gamma_{32}\hat{\sigma}_{32} & \Gamma_{43}\hat{\sigma}_{44}-\Gamma_{31}\hat{\sigma}_{33} & -\frac{1}{2}\gamma_{43}\hat{\sigma}_{34}\\
-\frac{1}{2}\gamma_{41}\hat{\sigma}_{41} & -\frac{1}{2}\gamma_{42}\hat{\sigma}_{42} & -\frac{1}{2}\gamma_{43}\hat{\sigma}_{43} & -(\Gamma_{42}+\Gamma_{43})\hat{\sigma}_{44}
\end{smallmatrix}
\end{bmatrix}, \label{eq:(A7)}
\end{align}
where the $ij$ matrix element represents $\hat{R}_{ij}$. $\Gamma_{ij}$ is the spontaneous decay rate from $\ket{i}$ to $\ket{j}$, and $\gamma_{ij}$ represents the decoherence rate between states $\ket{i}$ and $\ket{j}$. The specific form of the Langevin noise operator is not significant if the system-reservoir interaction can be considered as a Markovian process. The relation between the atomic system and the Langevin noise operator can be obtained from the Einstein relation, which will be further discussed in Appendix \ref{sec:Appendix B}.


\section{EINSTEIN RELATIONS AND DIFFUSION COEFFICIENTS} \label{sec:Appendix B}

Consider the system-reservoir interaction as a Markovian process, then the collective Langevin noise operators satisfy the following delta correlation \cite{Quantum Optics}:
\begin{align}
\langle\widetilde{F}_{ij}(z,\omega)&\widetilde{F}_{mn}(z',\omega')\rangle\notag\\
&=\frac{L}{2\pi N}D_{ij,mn}(z)\delta(z-z')\delta(\omega+\omega'), \label{eq:(B1)}
\end{align}
where $D_{ij,mn}(z)$ represents the diffusion coefficient, which is connected to the atomic system through the following Einstein relation \cite{Quantum Optics}:
\begin{align}
D_{ij,mn}&=\delta_{jm}\braket{\hat{R}_{in}}-N_z\braket{\hat{R}_{ij}\hat{\sigma}_{mn}}-N_z\braket{\hat{\sigma}_{ij}\hat{R}_{mn}}. \label{eq:(B2)}
\end{align}
Here, $N_z=\frac{N}{L}\Delta_z$ denotes the number of atoms within $\Delta_z$ around position $z$; the relaxation term can be expanded as
\begin{align}
\hat{R}_{ij}=\sum_{mn}\Gamma_{ij,mn}\hat{\sigma}_{mn}, \label{eq:(B3)}
\end{align}
where $\Gamma_{ij,mn}$ characterizes the spontaneous decay rate and decoherence rate of the relaxation process. By using the property $\hat{\sigma}_{ij}\hat{\sigma}_{kl}=\frac{1}{N_z}\delta_{jk}\hat{\sigma}_{il}$, the diffusion coefficients can be obtained without deriving the explicit form of the Langevin noise operators. Since Markovian approximation is reasonable for the diamond-type QFC system, the normal-order diffusion coefficients $D_{\alpha_i^{\dagger},\alpha_j}$ considered in the main text are as follows:
\begin{align}
D_{\alpha_i^{\dagger},\alpha_j}=&\begin{bmatrix}
D_{21,12} & D_{21,14} & D_{21,32} & D_{21,34}\\
D_{41,12} & D_{41,14} & D_{41,32} & D_{41,34}\\
D_{23,12} & D_{23,14} & D_{23,32} & D_{23,34}\\
D_{43,12} & D_{43,14} & D_{43,32} & D_{43,34}
\end{bmatrix}\notag\\
=&\begin{bmatrix}
D_{21,12} & D_{21,14} & 0 & 0\\
D_{41,12} & D_{41,14} & 0 & 0\\
0 & 0 & D_{23,32} & D_{23,34}\\
0 & 0 & D_{43,32} & D_{43,34}
\end{bmatrix}. \label{eq:(B4)}
\end{align}
The remaining terms consist of linear combinations involving $\braket{\hat{\sigma}_{22}}$, $\braket{\hat{\sigma}_{24}}$, $\braket{\hat{\sigma}_{42}}$, and $\braket{\hat{\sigma}_{44}}$. In the main text, the delta correlation should take the form specified in Eq. (\ref{eq:(19)}) under steady-state conditions. Additionally, we assume that $\hat{\sigma}_{ij}\approx\hat{\sigma}_{ij}^{(0)}+\hat{\sigma}_{ij}^{(1)}$ during the calculation of the diffusion coefficients. The relevant atomic operators are as follows:
\begin{align}
\hat{\sigma}_{22(24,42,44)}&\approx\hat{\sigma}_{22(24,42,44)}^{(0)}+\hat{\sigma}_{22(24,42,44)}^{(1)}\notag\\
&=\sum_{ij}K_{ij}^{22(24,42,44)}\widetilde{F}_{ij}, \label{eq:(B5)}
\end{align}
where we employ the result $\braket{\hat{\sigma}_{22(24,42,44)}^{(0)}}=0$. By using the property $\braket{\widetilde{F}_{ij}}=0$ for the the vacuum reservoir, we find that all remaining diffusion coefficients are zero. For the subspace discussed in the main text, this implies $D_{\alpha_i^{\dagger},\alpha_j}=0$.


\section{COUPLING MSE SOLUTION} \label{sec:Appendix C}

Consider the steady-state solution for the coupling field MSE, in which the coupling field has reached a stable state, and the Rabi frequency remains constant over time. The coupling field MSE can then be simplified into the following form:
\begin{align}
\frac{\partial}{\partial z}\Omega_c(z)=\frac{i \alpha_c \Gamma_{31}}{2 L}\braket{\hat{\sigma}_{13}^{(0)}(z)}=\frac{C_0 \Omega_c}{A_0+B_0\Omega_c \Omega_c^*}, \label{eq:(C1)}
\end{align}
where we define $A_0=2L\Gamma_{31}(\gamma_{31}^2+4\Delta_c^2)$, $B_0=4L\gamma_{31}$, and $C_0=-\alpha_c\Gamma_{31}^2(\gamma_{31}+2 i \Delta_c)$. If we let $u=\Omega_c \Omega_c^*$, solving the corresponding differential equation for $u(z)$ yields the exact solution as follows:
\begin{align}
&u(z)=\frac{A_0}{B_0}W_0\left[\frac{B_0}{A_0}u(0)e^{\frac{D_0}{A_0}z+\frac{B_0}{A_0}u(0)}\right]. \label{eq:(C2)}
\end{align}
Here, we define $D_0=C_0+C_0^*$, and $W_0(x)$ represents the principal branch of the Lambert W function \cite{Lambert W Function}. Upon substituting this solution back into the coupling field MSE, the resulting differential equation is
\begin{align}
\frac{\partial}{\partial z}\Omega_c(z)-G_0(z)\Omega_c(z)=0, \label{eq:(C3)}
\end{align}
where $G_0=\frac{C_0}{A_0+B_0u(z)}$, and the differential equation can be solved directly by utilizing the integrating factor:
\begin{align}
&\Omega_c(z)=\Omega_c(0)e^{\int_{0}^{z} G_0(z) dz}. \label{eq:(C4)}
\end{align}
This represents the exact solution to the coupling field MSE, elucidating the attenuation of the coupling field as it propagates through the atomic medium.


\section{COEFFICIENTS OF COUPLED EQUATIONS} \label{sec:Appendix D}

The self-coupling and cross-coupling coefficients, derived from the coupled equations (\ref{eq:(13)}) and (\ref{eq:(14)}), manifest as follows:
\begin{align}
\Lambda_p&(z,\omega)\notag\\
=&\frac{2iN|g_p|^2}{cT_0}\bigg[\braket{\hat{\sigma}_{31}^{(0)}}\gamma_{41}'\gamma_{43}'\Omega_c\left(1+\frac{|\Omega_c|^2-|\Omega_d|^2}{\gamma_{41}'\gamma_{43}'}\right)\notag\\
+&\braket{\hat{\sigma}_{11}^{(0)}}i\gamma_{32}'\gamma_{41}'\gamma_{43}'\left(1+\frac{|\Omega_c|^2}{\gamma_{41}'\gamma_{43}'}+\frac{|\Omega_d|^2}{\gamma_{32}'\gamma_{43}'}\right)\bigg]+\frac{i\omega}{c}, \label{eq:(D1)}\\
\kappa_p&(z,\omega)\notag\\
=&\frac{2iNg_p^*g_s}{cT_0}e^{-i\Delta k z}\bigg[\braket{\hat{\sigma}_{13}^{(0)}}\gamma_{32}'\gamma_{43}'\Omega_d^*\left(\frac{|\Omega_c|^2-|\Omega_d|^2-1}{\gamma_{32}'\gamma_{43}'}\right)\notag\\
+&\braket{\hat{\sigma}_{33}^{(0)}}i(\gamma_{32}'+\gamma_{41}')\Omega_c\Omega_d^*\bigg], \label{eq:(D2)}\\
\Lambda_s&(z,\omega)\notag\\
=&\frac{2iN|g_s|^2}{cT_0}e^{-i\Delta k z}\bigg[\braket{\hat{\sigma}_{13}^{(0)}}\gamma_{21}'\gamma_{32}'\Omega_c^*\left(1+\frac{|\Omega_c|^2-|\Omega_d|^2}{\gamma_{21}'\gamma_{32}'}\right)\notag\\
+&\braket{\hat{\sigma}_{33}^{(0)}}i\gamma_{21}'\gamma_{32}'\gamma_{41}'\left(1+\frac{|\Omega_c|^2}{\gamma_{21}'\gamma_{32}'}+\frac{|\Omega_d|^2}{\gamma_{21}'\gamma_{41}'}\right)\bigg]+\frac{i\omega}{c}, \label{eq:(D3)}
\end{align}
\begin{align}
\kappa_s&(z,\omega)\notag\\
=&\frac{2iNg_s^*g_p}{cT_0}\bigg[\braket{\hat{\sigma}_{31}^{(0)}}\gamma_{21}'\gamma_{41}'\Omega_d\left(\frac{|\Omega_c|^2-|\Omega_d|^2-1}{\gamma_{21}'\gamma_{41}'}\right)\notag\\
+&\braket{\hat{\sigma}_{11}^{(0)}}i(\gamma_{32}'+\gamma_{41}')\Omega_c^*\Omega_d\bigg], \label{eq:(D4)}
\end{align}
where $\gamma_{21}'=\gamma_{21}-2i(\Delta_p+\omega)$, $\gamma_{32}'=\gamma_{32}-2i(\Delta_p-\Delta_c+\omega)$, $\gamma_{41}'=\gamma_{41}-2i(\delta+\omega)$, $\gamma_{43}'=\gamma_{43}-2i(\delta-\Delta_c+\omega)$, and $T_0$ is defined as follows:
\begin{align}
T_0=&|\Omega_c|^2(\gamma_{21}'\gamma_{32}'+\gamma_{41}'\gamma_{43}')+|\Omega_d|^2(\gamma_{21}'\gamma_{41}'+\gamma_{32}'\gamma_{43}')\notag\\
+&\left(|\Omega_c|^2-|\Omega_d|^2\right)^2. \label{eq:(D5)}
\end{align}
The coefficients of noise disturbance are not listed here because, throughout all the discussions in this paper, it has been demonstrated that all terms associated with these coefficients are zero. This is due to the determination of the related diffusion coefficients being zero, as calculated from the corresponding Einstein relations.


\section{MAGNUS EXPANSION} \label{sec:Appendix E}

We begin by considering a first-order linear ODE for the two-dimensional linear operator, with the initial condition at $z=0$. The ODE can be expressed as follows:
\begin{align}
&\frac{\partial}{\partial z}X(z,\omega)=M(z,\omega)X(z,\omega), \label{eq:(E1)}
\end{align}
where $X(z,\omega)$ and $M(z,\omega)$ are defined as follows:
\begin{align}
X(z,\omega)&=\begin{bmatrix}
\widetilde{a}_p(z,\omega)\\
\widetilde{a}_s(z,\omega)
\end{bmatrix}, \label{eq:(E2)}\\
M(z,\omega)&=\begin{bmatrix}
\Lambda_p(z,\omega) & \kappa_p(z,\omega)\\
\kappa_s(z,\omega) & \Lambda_s(z,\omega)
\end{bmatrix}. \label{eq:(E3)}
\end{align}
The general solution to the first-order homogeneous linear ODE was proposed by Magnus \cite{Magnus Expansion}, where he presents an exponential solution for the linear operator. Thus, the solution to Eq. (\ref{eq:(E1)}) can be expressed as
\begin{align}
&X(z,\omega)=e^{\Omega(z,\omega)}X(0,\omega), \label{eq:(E4)}
\end{align}
where $\Omega(z,\omega)$ is constructed as the following series expansion:
\begin{align}
&\Omega(z,\omega)=\sum_{n=1}^{\infty}\Omega_n(z,\omega),\label{eq:(E5)}\\
&\Omega_1(z,\omega)=\int_{0}^{z} M(s,\omega) ds. \label{eq:(E6)}
\end{align}
This formulation is known as the Magnus expansion \cite{Magnus Expansion Applications}. The higher-order terms of the Magnus series can be recursively obtained using the following relation \cite{Magnus Expansion Recursive}:
\begin{align}
&\Omega_n(z,\omega)=\sum_{k=1}^{n-1}\frac{B_k}{k!}\int_{0}^{z} ds S_n^{(k)}(s,\omega), \label{eq:(E7)}\\
&S_n^1=[\Omega_{n-1},M], \label{eq:(E8)}\\
&S_n^{(k)}=\sum_{m=1}^{n-k}[\Omega_m,S_{n-m}^{(k-1)}], \label{eq:(E9)}
\end{align}
where $B_k$ represents the $k$th Bernoulli number. It is noteworthy that if $M$ does not depend on $z$, the only remaining term in the Magnus series is $\Omega_1(z,\omega)$. In this scenario, the solution for $X(z, \omega)$ is equivalent to the one obtained using the integrating factor.

The linear ODE in the main text, encompassing Eqs. (\ref{eq:(13)}) and (\ref{eq:(14)}), can be expressed as follows:
\begin{align}
\frac{\partial}{\partial z}X(z,\omega)=M(z,\omega)X(z,\omega)+F(z,\omega), \label{eq:(E10)}
\end{align}
where we define $F(z,\omega)$ as follows:
\begin{align}
F(z,\omega)=\sum_{\alpha_i}
\begin{bmatrix}
\xi_{\alpha_i}^{p}(z,\omega)\\
\xi_{\alpha_i}^{s}(z,\omega)
\end{bmatrix}
\widetilde{f}_{\alpha_i}(z,\omega). \label{eq:(E11)}
\end{align}
Applying Eq. (\ref{eq:(E1)}) allows us to obtain the solution to Eq. (\ref{eq:(E10)}) as follows:
\begin{align}
X(z,\omega)=e^{\Omega(z,\omega)}X(0,\omega)+e^{\Omega(z,\omega)}\int_0^z e^{-\Omega(s,\omega)}F(s,\omega) ds. \label{eq:(E12)}
\end{align}
Through diagonalizing the $2 \times 2$ matrix $\Omega(z,\omega)$ and substituting the diagonalized result into the Taylor expansion of the exponential function $e^{\Omega(z,\omega)}$, the matrix form in Eq. (\ref{eq:(16)}) can be explicitly derived.


\section{KEY COMPONENTS IN QUANTUM STATE DERIVATION} \label{sec:Appendix F}

The outer product of the vacuum state $\ket{0}$ is directly linked to the ladder operator of the optical field \cite{Vacuum Outer Product Property}. This relationship for a single-mode field can be derived as follows:
\begin{align}
&\sum_{l=0}^{\infty}\frac{(-1)^l}{l!}(\widetilde{a}^{\dagger})^l(\widetilde{a})^l\notag\\
&=\sum_{l=0}^{\infty}\sum_{i,j=0}^{\infty}\frac{(-1)^l}{l!}\ket{i}\bra{j}\bra{i}(\widetilde{a}^{\dagger})^l(\widetilde{a})^l\ket{j}\notag\\
&=\sum_{l=0}^{\infty}\frac{(-1)^l}{l!}\sum_{i=l}^{\infty}\ket{i}\bra{i}\frac{i!}{(i-l)!}\notag\\
&=\sum_{i=0}^{\infty}\ket{i}\bra{i}\sum_{l=0}^{i-1}(-1)^l C^i_l+(-1)^i\notag\\
&=\sum_{i=1}^{\infty}\ket{i}\bra{i}\left[(-1)^i+(-1)^{i-1}C^{i-1}_{i-1}\right]+\ket{0}\bra{0}\notag\\
&=\ket{0}\bra{0}, \label{eq:(F1)}
\end{align}
which allows the outer product of two number states to be expressed as the sum of multiplication of ladder operators. 

To express $\braket{[\widetilde{a}_{s}^{\dagger}(L)]^{l+n}[\widetilde{a}_{s}(L)]^{l+m}}$ in a computationally friendly form, we need to substitute the explicit forms of the ladder operators, as given in the single-mode version of Eq. (\ref{eq:(17)}). Afterward, we expand the multiplication of the sum of the initial ladder operators and noise operators. Each term in the expansion takes the following form:
\begin{align}
\Bigg\langle\left[C^*(0)\widetilde{a}_{p}^{\dagger}(0)+\right.D^*(0)&\left.\widetilde{a}_{s}^{\dagger}(0)\right]^a\notag\\
&\left[\sum_{\alpha_i}\int_{0}^{L} dz Q_{\alpha_i}^*(z)\widetilde{f}_{\alpha_i}^{\dagger}(z)\right]^b\notag\\
\left[C(0)\widetilde{a}_{p}(0)\right.+D(0)&\left.\widetilde{a}_{s}(0)\right]^c\notag\\
&\left[\sum_{\alpha_i}\int_{0}^{L} dz Q_{\alpha_i}(z)\widetilde{f}_{\alpha_i}(z)\right]^d\Bigg\rangle, \label{eq:(F2)}
\end{align}
where $a,b,c,d\in \{0\cup\mathbb{N}\}$. These expanded terms can be classified into two categories: $b+d\neq 0$ or $b+d=0$. In the first case, the partial trace of the reservoir applies only to the noise operator parts, while the partial trace of the probe and signal applies to the remaining parts. The expansion of the noise part is as follows:
\begin{align}
\Biggl\langle\bigg[\sum_{\alpha_i}\int_{0}^{L} dz Q_{\alpha_i}^*(z)&\widetilde{f}_{\alpha_i}^{\dagger}(z)\bigg]^b\notag\\
&\bigg[\sum_{\alpha_i}\int_{0}^{L} dz Q_{\alpha_i}(z)\widetilde{f}_{\alpha_i}(z)\bigg]^d\Biggr\rangle_R&\notag\\
=\sum_{r}K_r\int_0^L dz_1 Q_{\alpha_{r,1}}^*&(z_1)\ldots \int_0^L dz_{b+d} Q_{\alpha_{r,b+d}}^*(z_{b+d})\notag\\
\langle\widetilde{f}_{\alpha_{r,1}}^{\dagger}(z_1)\ldots\widetilde{f}_{\alpha_{r,b}}^{\dagger}(z_b)&\widetilde{f}_{\alpha_{r,b+1}}(z_{b+1})\ldots\widetilde{f}_{\alpha_{r,b+d}}(z_{b+d})\rangle_R, \label{eq:(F3)}
\end{align}
where $\alpha_{r, j}$ represents an element of $\{12,14,32,34\}$. To simplify Eq. (\ref{eq:(F3)}), we apply the following generalized Wick's theorem \cite{Vacuum Outer Product Property, Wick's Theorem}:
\begin{align}
&\braket{\widetilde{F}_1\widetilde{F}_2\ldots\widetilde{F}_{2n}}_R\notag\\
&=\braket{\widetilde{F}_1\widetilde{F}_2}_R\braket{\widetilde{F}_3\widetilde{F}_4\ldots\widetilde{F}_{2n}}_R+\braket{\widetilde{F}_1\widetilde{F}_3}_R\braket{\widetilde{F}_2\widetilde{F}_4\ldots\widetilde{F}_{2n}}_R\notag\\
&+\ldots+\braket{\widetilde{F}_1\widetilde{F}_{2n}}_R\braket{\widetilde{F}_2\widetilde{F}_3\ldots\widetilde{F}_{2n-1}}_R, \label{eq:(F4)}
\end{align}
where $\widetilde{F}_i$ denotes the Langevin noise operator. Note that if the braket in Eq. (\ref{eq:(F4)}) includes an odd number of noise operators, the expectation value is zero. For $b+d\in {\rm even}$, each term in Eq. (\ref{eq:(F3)}) can be expanded as the sum of multiplication of the second-order correlation functions of the noise operators. By using the Einstein relations, $D_{\alpha_i,\alpha_j}$, $D_{\alpha_i^{\dagger},\alpha_j^{\dagger}}$, and $D_{\alpha_i^{\dagger},\alpha_j}$ are all zero; thus, all possible correlation functions are zero. As Eq. (\ref{eq:(F3)}) evaluates to zero, the only non-zero term in the expansion of $\braket{[\widetilde{a}_{s}^{\dagger}(L)]^{l+n}[\widetilde{a}_{s}(L)]^{l+m}}$ is as follows:
\begin{align}
\left\langle[C^*(0)\widetilde{a}_{p}^{\dagger}(0)\right.+D^*(0)&\widetilde{a}_{s}^{\dagger}(0)]^{l+n}\notag\\
&\left.[C(0)\widetilde{a}_{p}(0)+D(0)\widetilde{a}_{s}(0)]^{l+m}\right\rangle. \label{eq:(F5)}
\end{align}
The partial trace applied to the reservoir density operator equals 1. Therefore, we only apply the partial trace to the probe and signal in Eq. (\ref{eq:(F5)}). We can expand Eq. (\ref{eq:(F5)}), and each term in the expansion takes the following form:
\begin{align}
\left\langle[C^*(0)\widetilde{a}_{p}^{\dagger}(0)]^a\right.[D^*(0)&\widetilde{a}_{s}^{\dagger}(0)]^b\notag\\
&\left.[C(0)\widetilde{a}_{p}(0)]^c[D(0)\widetilde{a}_{s}(0)]^d\right\rangle_{p,s}. \label{eq:(F6)}
\end{align}
These terms can again be classified into two categories with $b+d\neq 0$ or $b+d=0$. For the first case, the partial trace on the signal ladder operator part is zero since the signal field is initially in a vacuum state $\ket{0}_s$. Therefore, the only non-zero term in the expansion of Eq. (\ref{eq:(F5)}) is the $b+d=0$ term, for which only the probe ladder operator part remains. The relation that we use in Eq. (\ref{eq:(55)}) can be obtained as follows:
\begin{align}
&\left\langle[\widetilde{a}_{s}^{\dagger}(L)]^{l+n}[\widetilde{a}_{s}(L)]^{l+m}\right\rangle\notag\\
&=tr_p\left\{[C^*(0)\widetilde{a}_{p}^{\dagger}(0)]^{l+n}[C(0)\widetilde{a}_{p}(0)]^{l+m}\rho_p(0)\right\}. \label{eq:(F7)}
\end{align}
The explicit form of the converted density matrix element can then be derived as shown in Eq. (\ref{eq:(55)}).



\end{document}